\documentclass[reprint,aps, twocolumn, superscriptaddress]{revtex4}
\usepackage[normalem]{ulem}
\usepackage{graphicx}% Include figure files
\usepackage{dcolumn}% Align table columns on decimal point
\usepackage[mathlines]{lineno}
\usepackage{hyperref}
\usepackage{bm}% bold math
\usepackage{amssymb}
\usepackage{amsmath}
\usepackage{braket}
\usepackage{color}
\usepackage{xcolor}
\usepackage{soul} 
\usepackage{graphics}
\usepackage{bbm}

\begin{document}
\title{A dimensionality and purity measure for high-dimensional entangled states}
\author{Isaac Nape} 
\affiliation{School of Physics, University of the Witwatersrand, Private Bag 3, Wits 2050, South Africa}
\author{ Valeria Rodr\'iguez-Fajardo}
\affiliation{School of Physics, University of the Witwatersrand, Private Bag 3, Wits 2050, South Africa}
\author{Feng Zhu} 
\affiliation{School of Engineering and Physical Sciences, Heriot-Watt University, Edinburgh, EH14 4AS, UK}
\author{Hsiao-Chih Huang} 
\affiliation{Department of Physics, National Taiwan University, Taipei 106, Taiwan}
\author{Jonathan Leach} 
\affiliation{School of Engineering and Physical Sciences, Heriot-Watt University, Edinburgh, EH14 4AS, UK}
\author{Andrew Forbes}
\affiliation{School of Physics, University of the Witwatersrand, Private Bag 3, Wits 2050, South Africa}

%%%%%%%%%%%%%%%%%%%%%%%%%%%%%%%%%%%%%%%%%%%%%%%%%%
%%% Abstract
%%%%%%%%%%%%%%%%%%%%%%%%%%%%%%%%%%%%%%%%%%%%%%%%%%
\begin{abstract}
\noindent High-dimensional entangled states are promising candidates for increasing the security and encoding capacity of quantum systems.  While it is possible to witness and set bounds for the entanglement, precisely quantifying the dimensionality and purity in a fast and accurate manner remains an open challenge. Here, we report an approach that simultaneously returns the dimensionality and purity of high-dimensional entangled states by simple projective measurements.  %We achieve this by engineering superpositions of fractional OAM modes such that the resulting azimuthal phase has $n$ identical discontinuities.  
We show that the outcome of a conditional measurement returns a visibility that scales monotonically with entanglement dimensionality and purity, allowing for quantitative measurements for general photonic quantum systems. We illustrate our method using transverse spatial modes of photons that carry orbital angular momentum and verify high-dimensional entanglement over a wide range of state purities. Our approach advances the high-dimensional tool box for characterising quantum states by providing a simple and direct dimensionality and purity measure, even for mixed entangled states.
%In practise, such  quantum systems are generally represented as mixed states, whose purity can be degraded by various noise mechanisms.
\end{abstract}
\maketitle

%%%%%%%%%%%%%%%%%%%%%%%%%%%%%%%%%%%%%%%%%%%%%%%%%%
%%% Introduction
%%%%%%%%%%%%%%%%%%%%%%%%%%%%%%%%%%%%%%%%%%%%%%%%%%
\noindent High-dimensional entangled states are widely used throughout quantum science to increase secure information bandwidth and security bounds for quantum communication \cite{cozzolino2019high}. Through the precise control of high dimensional photonic states \cite{erhard2017quantum, babazadeh2017high}, i.e., time-energy, transverse momentum, spatial degrees of freedom or all of them simultaneously \cite{deng2017quantum}, the unparalleled benefits of high dimensional state encoding are taking center stage. Recent developments in this direction have displayed the feasibility of quantum information processing that is robustness against optimal quantum cloning machines \cite{gisin1997optimal, bouchard2017high}, environmental noise \cite{ecker2019overcoming} and improved information rates \cite{dixon2008gigahertz, barreiro2008beating}, demonstrating a significant advantage in comparison to traditional qubit encoding.

Despite the advantages of high-dimensional quantum states, certifying and quantifying the dimensionality of such systems still remains challenging, particularly in the presence of noise.  The intuitive approach of simply measuring the width of the modal spectrum is a necessary but not sufficient condition to determine dimensionality as it fails to account for non-local correlations. Consequently, many techniques have been developed to witness, bound and attempt to quantify high-dimensional quantum states. These include approximating the density matrix via quantum state tomography (QST) with multiple qubit state projections \cite{Agnew2011}, using mutually unbiased bases \cite{giovannini2013characterization,bavaresco2018measurements} to probe the states, and testing non-local bi-photon correlations by generalised Bell tests in higher dimensions \cite{vaziri2002experimental, groblacher2006experimental, dada2011experimental,oemrawsingh2005experimental, oemrawsingh2006high,gotte2007quantum,huang2018various}.  
However, the spectrum measurements do not confirm entanglement, the QST approach scales unfavourably with dimension, only bounds or witnesses are possible with the mutually unbiased bases method and the dimension to be probed must be known \textit{a priori} (e.g., valid for prime or prime power dimensions), and finally, the high-dimensional Bell tests can fail the fair sampling condition \cite{dada2011bell, romero2013tailored}.  A further limitation in the present state-of-the-art is that certain dimensionality measurements consider only pure states \cite{pors2008shannon, giovannini2013characterization}, yet noise mechanisms always introduce some degree of mixture to the system. This has a detrimental effect on the accuracy of measured dimensions due to the reduced purity \cite{zhu2019high}. Consequently, no approach allows both the purity and dimensionality of arbitrary high dimensional mixed states to be quantitatively deduced in a simple and accurate manner. Yet, knowing the purity and dimension of the state is crucial for fundamental tests of quantum mechanics as well as for quantum information processing protocols, setting the required violation of inequalities in the former, and the information capacity of the state, the allowed error bounds in secure communication systems, and the requirement for entanglement distillation in the latter.

In this work we will provide a solution to this pressing problem and illustrate it using the topical example of transverse spatial modes of photons carrying orbital angular momentum (OAM), which have been instrumental in realising quantum entanglement beyond qubits \cite{fickler2012quantum, krenn2015twisted, erhard2017quantum,erhard2018twisted, fickler2014interface, forbes2019quantum, molina2007twisted}. They have emerged as an ideal resource in quantum information processing and communication, including superdense coding \cite{wang2005quantum, barreiro2008beating}, multi-photon entanglement \cite{Malik2016}, quantum teleportation \cite{goyal2014qudit} and entanglement swapping \cite{zhang2017simultaneous,bornman2019ghost}, ghost imaging \cite{jack2009holographic,chen2014quantum,bornman2019ghostaa} and secure communication in free-space \cite{mafu2013higher,steinlechner2017distribution} and optical fibre \cite{liu2020multidimensional,cao2020distribution,cozzolino2019air,cozzolino2019orbital}, fuelled by their inherent advantages such as robustness against noise \cite{ecker2019overcoming} and higher information capacity per photon \cite{leach2012secure}.  Further, the toolkit to engineer high-dimensional states is readily available, e.g., by linear \cite{zhang2016engineering} and non-linear processes \cite{torres2003quantum, miatto2011full, Romero2012,Terriza, zhang2014simulating}, with up to 100$\times$100 dimensions already demonstrated \cite{krenn2014generation}. 

With OAM as our example, we present a scheme to simultaneously quantify the dimensionality and purity of a two-particle entangled state. By measuring coincidence fringes from projective measurements using analysers acting on the entangled photons, we are able to accurately measure the dimensionality and purity of our entangled state from the visibility, which is only reproducible by entangled photons. We first outline the concept and theory and then demonstrate it experimentally on states with arbitrary purity. Our quantitative technique is simple, fast and robust, making it ideal for practical implementations (even with undesired noise) of quantum protocols with general high-dimensional photonic quantum entangled states.

%%%%%%%%%%%%%%%%%%%%%%%%%%%%%%%%%%%%%%%%%%%%%%%%%%
\section{Concept}
\begin{figure*}[t!]
\centering
\includegraphics[width=\linewidth]{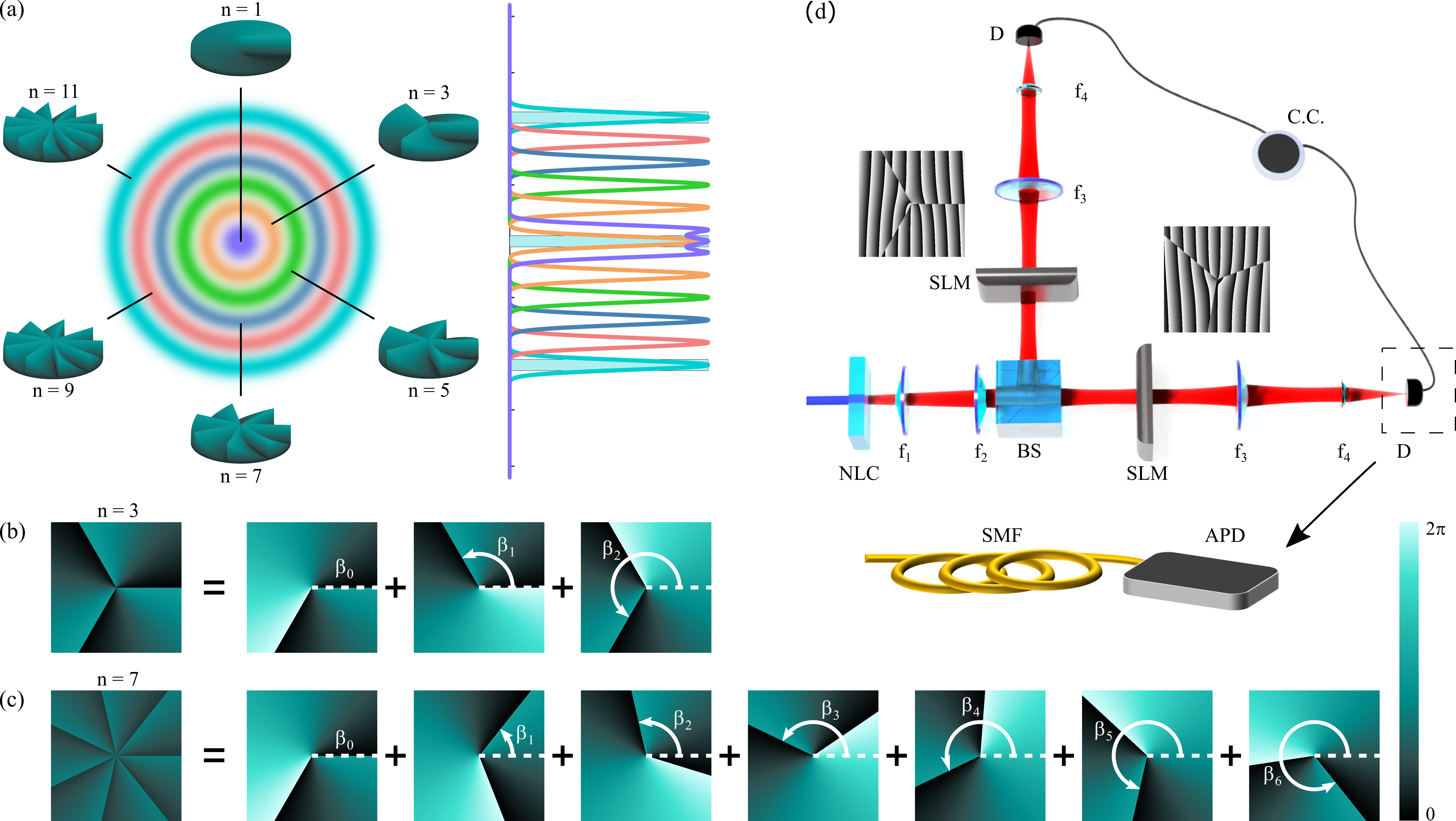}
\caption{(a) Conceptual visualisation of different analysers sampling various portions of a high dimensional discrete Hilbert space. Mode analysers construction for (b) $n=3$ and (c) $n=7$ superpositions of fractional OAM states.  (d) Schematic of the experimental setup used to measure the dimensionality and purity of a quantum system. (NLC: Nonlinear crystal, f$_{1,2,3,4}$: lens, BS: 50:50 Beam-splitter, SLM: Spatial light modulator, D: detector, APD: avalanche photo diode coupled to a single mode fiber (SMF), C.C.: coincidence counter.)}
\label{fig:ConceptFig}
\end{figure*}
%%%%%%%%%%

The task here is to measure the effective dimensions and purity of an entangled system.  In general, the purity of the quantum system, and therefore the entanglement between photon pairs, is reduced due to noise introduced by the environment or detection system, usually background white noise produced by ambient light or high dark counts in single photon detectors or unwanted multiphoton events \cite{zhu2019high,ecker2019overcoming}. Such noisy quantum systems can be modelled by an isotropic state following
%%% Equation
\begin{equation}
   \rho = p \ket{\Psi}\bra{\Psi} + \left( \tfrac{1-p}{d^2} \right) \mathbbm{I}_{d^2},
   \label{eq:isostate}
\end{equation}
%%%%%%
\noindent which considers contributions of both pure, $\ket{\Psi}$, and mixed, $\mathbbm{I}_d^2$ (d$^2$-dimensional identity operator), parts. Here $p$ is a parameter that determines the purity of the isotropic state, and varies from a maximally mixed (separable) state for $p=0$ to a completely pure (entangled) state for $p=1$. The dimension of the pure part of the state $\rho$ can be quantified by the Schmidt number, $K= 1/\sum_{j}|\lambda_j|^2$, where $\lambda_j$ are the Schmidt coefficients obtained from the decomposition of $\ket{\Psi}$ onto the basis states $\ket{\zeta_j}\in \mathcal{H}^{2}_\infty$, therefore spanning the two photon infinite dimensional Hilbert space.   We aim to find the dimensionality of the pure part of the system while also determining the purity assuming that the noise encroaching on the system is due to white noise from the environment.

The working principle of our technique is visualised in Fig. \ref{fig:ConceptFig}(a), where a set of analysers probe distinct parts of a discrete Hilbert space. We can think of each analyser as a probe that scans a sparse set of modes, reminiscent of a conditional measurement that indicates whether there is entanglement within the subspace or not. By combining the information gathered from a number of such analysers, we infer how many dimensions the state occupies. We will demonstrate this procedure both theoretically and experimentally.

To illustrate this concept, we consider a pair of entangled photons generated from spontaneous parametric down conversion (SPDC). Such photons can be entangled in their polarisation, energy-time, momentum or spatial mode \cite{forbes2019quantum} such as orbital angular momentum (OAM). Due to the great potential of the latter, particularly for quantum communications, we illustrate and demonstrate our method for OAM entangled states. In this case, the basis states, $\ket{\zeta_\ell}\rightarrow\ket{\ell}\ket{-\ell}$, are associated with an azimuthal phase profile $\text{exp}( i \ell \phi)$, with $\ell\in\mathbb{Z}$ being the topological charge and $\ell \hbar$ OAM per photon. An OAM entangled pure state can be expressed as
%%% equation
\begin{equation}
 \ket{\Psi}=\sum_{\ell=-L}^L \lambda_{\ell}\ket{\ell}_A \ket{-\ell}_B,
 \label{spdc}
\end{equation}
%%%
\noindent where $L$ is a positive integer denoting the largest mode index with a nonzero probability (simple to deduce by observing the counts), and $|\lambda_{\ell}|^2$ is the probability of generating photons in the states $\ket{\pm\ell}$ for photons $A$ and $B$, respectively.  While in general the state $\ket{\Psi}$ can be represented using an unbounded number of eigenmodes, i.e., $L\rightarrow \infty$, we truncate $\ket{\Psi}$ to the dimensions specified by $K$. That is, we choose $L=(K-1)/2$. This validly approximates the state in terms of its effective dimensions, enabling for its representation on a finite Hilbert space. %The distribution of $|\lambda_{\ell}|^2$ relates directly to the width of the OAM entangled spectrum, i.e. the Schmidt number, $K$, which is turn is associated with the dimensionality of $\ket{\Psi}$. For example, a $d$-dimensional maximally entangled state has an OAM spectrum that is uniformly distributed, $|\lambda_\ell|^2=d^{-1}=(2L+1)^{-1}$, resulting in $K=d$ (see Supplementary Information for further examples and details).

To gain access to various parts of the Hilbert space, we make use of high dimensional mode projectors that map onto the states 
%%% equation

\begin{equation}
\ket{M,\alpha}_n = \mathcal{N} \sum^{2L}_{j=0} c_{w_j,M}^n(\alpha) \ket{j},
\label{eq:NFracState2}
\end{equation}

%%%
\noindent where $\mathcal{N}$ is a normalisation factor, $\ket{j}$ are the basis states on the $d=2L+1$ dimensional space. The coefficients, $c^{n}_{w_j, M}(\alpha)$, control the amplitudes and phases of the modes in the superposition (see Methods section). For OAM basis states, the coefficients can be represented accordingly by replacing the index $w_j$ with the topological charge $\ell=j-L$. Examples of the phase profiles for two such analysers are shown in Fig.~\ref{fig:ConceptFig} (d) and (c) for $n=3$ and $n=7$, respectively, with full details on their construction in the Methods section and Supplementary Information.  While $n$ and $M$ can be chosen arbitrarily, we find it optimal to set $n$ as an odd positive integer and $M=n/2$. %\cite{huang2018various}

\begin{figure*}[t!]
\centering
\includegraphics[width=\linewidth]{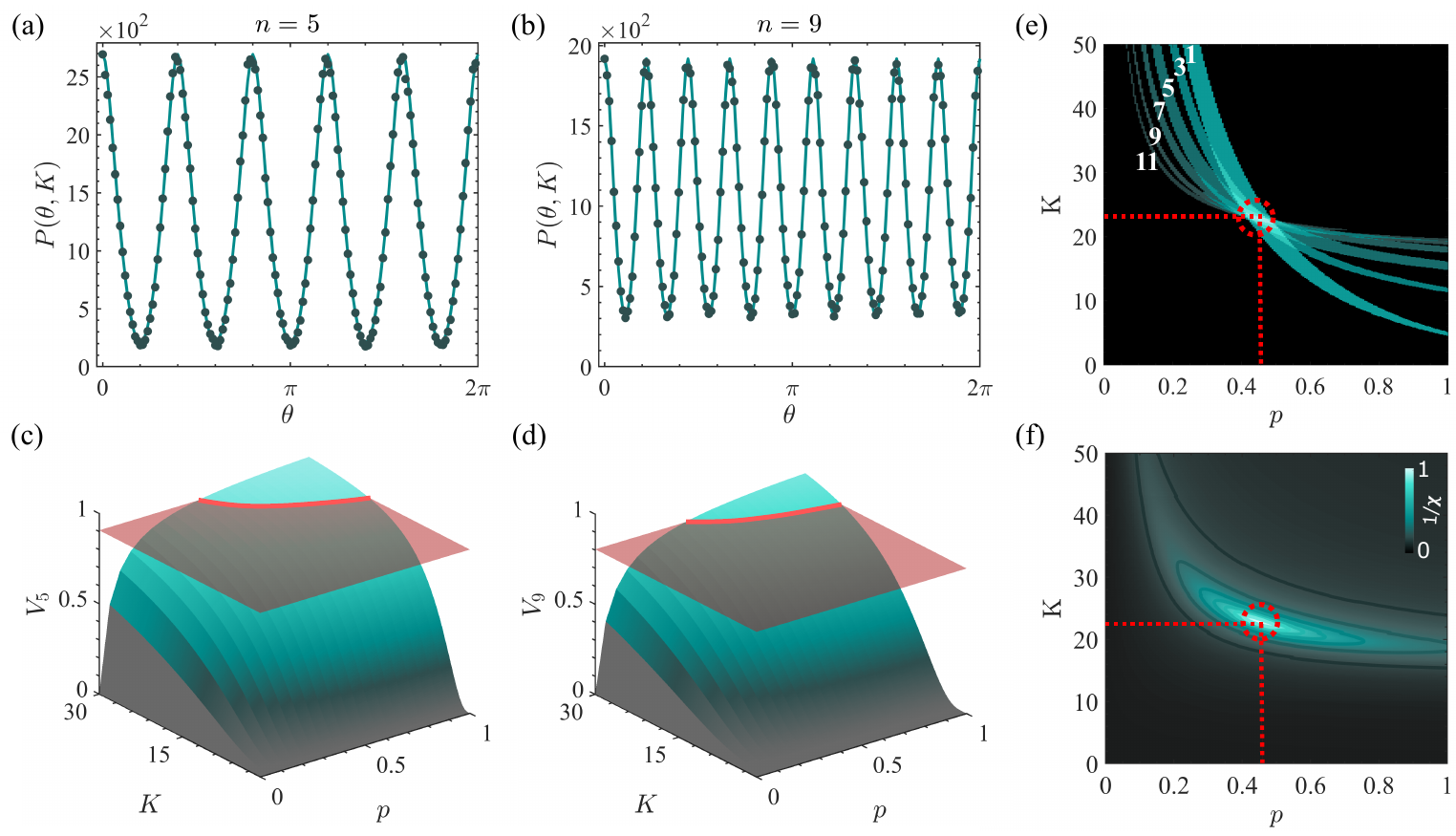}
\caption{Experimental (points) and theoretical (solid lines) coincidence count rates resulting from projections of photons $A$ and $B$ onto the states $\ket{M, \theta}_n$ and  $\ket{-M, 0}_n$, respectively, as a function of the relative orientation angle $\theta$ for (a) $n=5$ and  (b) $n=9$. Theoretical visibility as a function of the dimensionality ($K$) and purity ($p$) for (c) $n=5$ and (d) $n=9$, exemplifying it increases monotonically with both parameters. The (red) planes intersecting the curves are the experimental visibilities, with the possible solution space for each shown as a red trajectory.  The resulting trajectories for $n=1,2,...,11$ are shown in (e), where the thickness of each is due to the uncertainty in the visibility outcome. The dimension and purity of the system are found where they coincide, shown as a dashed red circle. The later corresponds to the optimal $(p,K)$ that minimise the function $\chi^2(p,K)$, or, equivalently, maximizes (e) $\sqrt{1/\chi^2}$, where the minimum of $\chi^2$ is now shown as a peak corresponding to $ (p, K)=(0.45 \pm 0.03, 22.84 \pm 0.62) $.}
% As the number superpositions $n$ increases so does sensitivity to the dimensions. 
\label{fig:result_fig}
\end{figure*}

Next, we project each photon in the isotropic state in Eq. (\ref{eq:isostate}), onto identical but conjugated analysers orientated by an angle $\theta$ relative to one another. A typical experimental setup for implementing this is sketched in Fig. \ref{fig:ConceptFig} (d). Entangled photon pairs are generated from SPDC and subsequently projected onto the states $\ket{M, \theta}$ and $\ket{-M, 0}$ by means of holograms having the transmission functions $U_{n}(\phi; \theta)$ and $U_{n}^{*}( \phi; 0)$,  respectively. In the OAM degree of freedom, the holograms correspond to fractional OAM modes \cite{gotte2007quantum}, known to have a non-integer azimuthal phase gradient. The modulated photons are then coupled into single mode fibers and measured in coincidences. The outcome probability of such a measurement, i.e., $\bra{ 0, -M}_n\bra{ \theta, M}_n \rho \ket{ M, \theta}_n\ket{ -M , 0}_n $, is 
%%% eq
\begin{equation}\label{eq:OVERLAPSPDC}
  P_n(\theta; p, K)  = p P_n(\theta, K) + \frac{1-p}{K^2} I_n(0, K),
\end{equation}
% %%% eq
% \begin{equation}\label{eq:OVERLAPSPDC}
% P_n(\theta) = \left | \sum_{\ell=-\infty}^{\infty} \lambda_{\ell}  \  (c^{n}_{\ell, M}(0) \ c^{n}_{-\ell,-M} (\theta))^* \right |^2,
% \end{equation} 
% %%
\noindent where $I_{n}(0, K)/K^2$ is the probability resulting from the overlap of the analysers with the maximally mixed state and $P_n(\theta, K)=\left|\sum_{\ell=-L}^{L}\lambda_{\ell}c^{n}_{\ell,M}(0)c^{n}_{-\ell,M}(\theta)\right|^2$ is the overlap probability with the pure state, with $M=n/2$ the fractional charge and $\lambda_{\ell}$ the initial bi-photon OAM spectrum. For a pure state, the probability curves have a parabolic shape following $P_n(\theta) = (\pi(2t-1)-n\theta)^2/\pi^2$, where $t=1,2,...\ n$. In Fig.~\ref{fig:result_fig} (a) and (b), we show as solid lines the theoretical probabilities (calculated using Eq.~(\ref{eq:OVERLAPSPDC})) of such a measurement as function of $\theta$. %Although not immediately obvious,  $p$ and $K$ can be inferred from the probability distribution in Eq. \ref{eq:OVERLAPSPDC} owing to its dependence on $\lambda_{\ell}$. 
We choose odd values of $n$ and $M=n/2$ to ensure a high visibility, which increases monotonically with $K$ and $p$ for each analyser (see Supplementary Information). In general both the shape and visibility of the fringes yield information about the state. %so that by a suitable minimisation procedure we can infer both the purity, $p$, and dimensionsality, $K$ (see Methods). 
To make the approach accurate and precise, we measure $N$ visibilities, $V_{n}$ for $n=1, 2, ..., \ 2N-1$, and infer the state properties by the intersection of their solution spaces. 
%%%%%%%%%%%%%%%%%%%%%%%%%%%%%%%%%%%%%%%%%%%%%%%%%%
\section{Results}
%%%%%%%%%%%%%%%%%%%%%%%%%%%%%%%%%%%%%%%%%%%%%%%%%%
%\noindent \textbf{Siral imaging and coindidence fringe measurements of downconversion photons.} 
The set-up used to demonstrate our scheme is shown conceptually in Fig.~\ref{fig:ConceptFig} (d) with the corresponding detailed description in the Methods section.  We measure the coincidences between the signal and idler photons for analyser projections on both arms as a function of the relative rotation angle of the holograms. To achieve this, we encoded the fractional OAM mode analyser on the SLM in the signal arm fixed at an angle $0$, while the conjugate mode was encoded in the idler arm and rotated at angles $\theta\in[0,2\pi]$.
% We measured high visibilities of $V=0.87 \pm 0.01$ and $V=0.72 \pm 0.01$.
%The observed fringes are consistent with the findings in \cite{oemrawsingh2005experimental} where a maximal visibility is observed when the fractional part is $\mu=0.5$.
%Initially, and in order to test that our analysers tailor the spiral spectrum as predicted in Eq.~(\ref{SPPnAmplitudes}), we projected one photon onto a fractional OAM (superposition) state and the other onto an OAM state. The results confirmed the validity of the holograms and are given in the Supplementary Information.  Next,
%%% figure
\begin{figure}[t]
\centering
\includegraphics[width=1\linewidth]{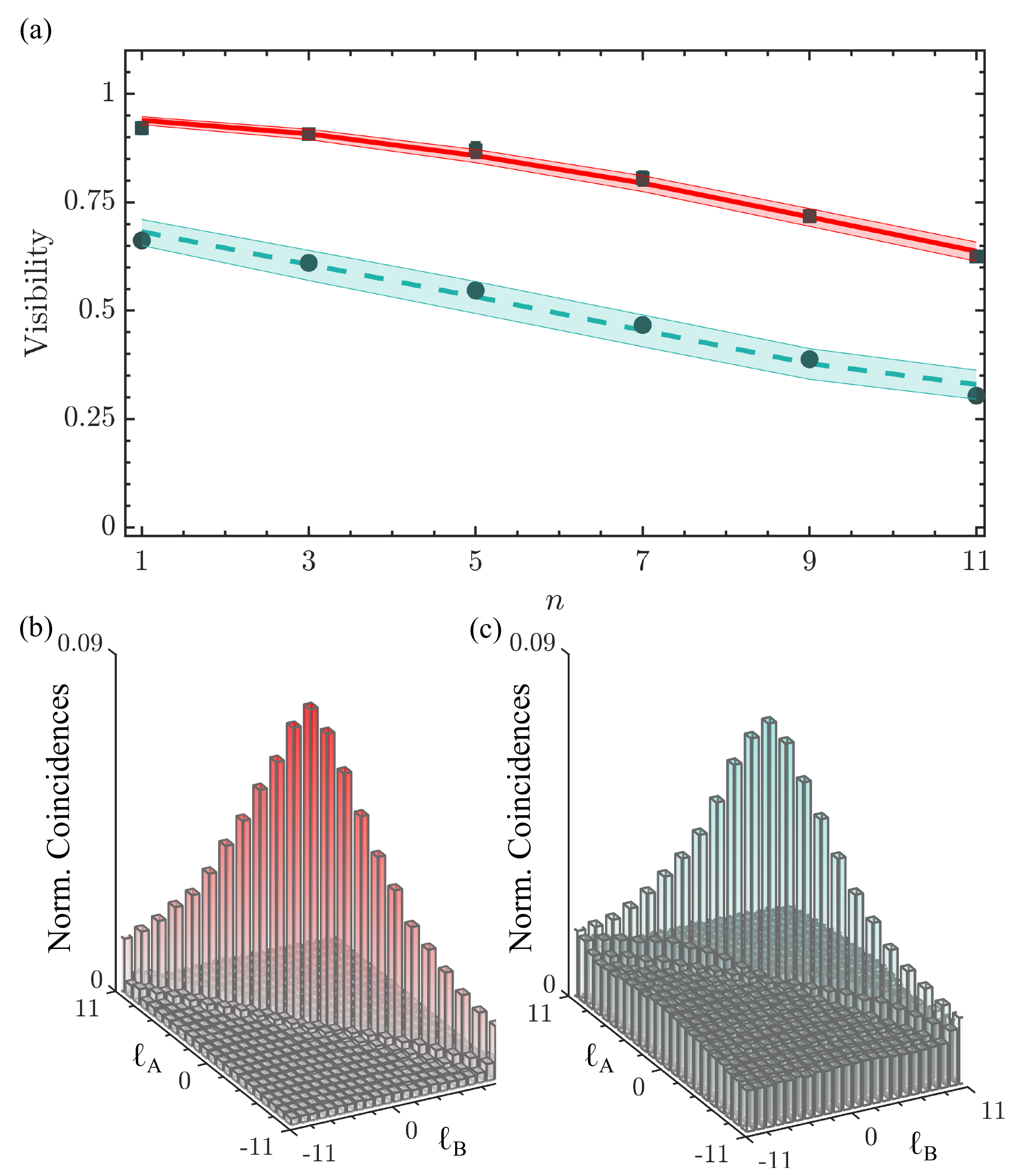}
\caption{Dimensionality and purity measurements assuming the SPDC mode spectrum for low (top) and high noise (bottom) levels. The points are the measured visibilities while the solid lines correspond to the fitted values. Measured spiral spectrum for the (b) low and (c) high noise levels.}
\label{fig:Visibility}
\end{figure}
%%%%%

%\noindent \textbf{Direct measurement of dimensionality from measured visibilities.}
To illustrate the operation of our technique we measured the coincidence-rates for six ($N=6$) analysers with $n=1,3,5,7,9$ and 11, and $M=n/2$, with example outcomes for $n=5$ and 9 shown as filled circles in Fig.~\ref{fig:result_fig}(a) and (b), respectively. Here, no accidental count subtraction was performed on the measurements. Importantly, the periodicity in the detected probabilities confirms the azimuthal $n$-fold symmetry predicted by our theory (solid curve).  Because the visibility is a monotonically increasing function of dimension and purity, a measured visibility returns a range of possible $(p,K)$ values, a ``trajectory'' or curve in the $(p,K)$ space.  This is illustrated in Fig.~\ref{fig:result_fig}(c) and (d), where the measured visibility (red horizontal plane) intercepts the visibility function along a curve (red curve) that restricts the possible solutions, $K$ and $p$, to those consistent with the measurement outcome.  The set of such curves from measuring many visibilities (each with its own analyser/projection) then restricts the final solution to a narrow region in $(p,K)$, whose uncertainty (width) is determined primarily from the uncertainly in the visibility measurement.  An example is shown in Fig.~\ref{fig:result_fig}(e), where each solution trajectory is projected onto the $(p,K)$ plane. Final values and uncertainty of $(p,K)$ can be determined by an appropriate routine to find the interception of all such trajectories by a minimisation procedure, as shown in Fig.~\ref{fig:result_fig}(f). %The final values and uncertainty in $(p,K)$ can be confirmed by an appropriate routine to find the interception of all such trajectories, illustrated by the red dashed circle, e.g., by a minimisation procedure, with the outcome for our example shown in Fig.~\ref{fig:result_fig}(f).
Using this approach we infer the purity and dimensionality of the system to be $(p, K) = (0.45\pm0.03, 22.84\pm0.62)$.  

In Fig.~\ref{fig:Visibility} (a) we show the six measured visibilities as square data points together with the calculated visibility (solid red line) based on the inferred $(p,K)$, which clearly match very well.  This is confirmation of the minimisation procedure for finding the intercept.  In order to assess the procedure under high noise levels, we introduced background noise using a white light source and repeated the measurements, shown as the circle data points and associated the blue dashed line in Fig.~\ref{fig:Visibility}. The average quantum contrast (see Supplementary Information), measured from the spiral spectrum in Fig. \ref{fig:Visibility} (c) and (d), dropped from $Q=19.19$ to $Q=3.76$, resulting in a reduced purity and dimensionality of $(p, K) = (0.13 \pm 0.01, 17.73\pm 0.71)$. %Following the sanity check of before, we find $p^\text{QC}=0.13$ (assuming $d = 19$) and a Schmidt number of $K^{SBW}=18$.

As a form of validation of these results, we estimate values from other techniques, with the comparison given in Table~\ref{tab:purity_dimen_summary}. If the dimension and noise of a system is known, then it is possible to calculate the purity following $\hat{p} = (Q-1)/(Q-1+d)$ where $Q$ is the quantum contrast and $d$ the dimension \cite{zhu2019high}. Likewise, if the state is assumed to be pure and not mixed and background subtraction is done to remove noise, then the spiral spectrum can be used to get an upper bound on the dimension.  For the two noise cases in Table~\ref{tab:purity_dimen_summary}, low and high, we find purity estimates of $\hat{p} \approx 0.44$ (assuming $d = 23$) and $\hat{p} \approx 0.13$ (assuming $d = 19$), respectively, while estimates of the dimensionality return $\hat{K} \approx 22$ and $\hat{K} \approx 18$, respectively. These values are in excellent agreement with our results, which did not require any such assumptions, nor any noise adjustments. The (perhaps) surprisingly low purity in our results is due to the fact that no background subtraction due to accidental counts was performed.
%%% Eq
%\begin{equation}\label{eq:purityZhu2019}
%\hat{p} = \frac{Q-1}{Q-1+d},
%\end{equation}
%%%%%
%%%%%%%%%%%table
\begin{table}[h!]
\centering
\begin{tabular}{|c|c|c|c|c|c|} \hline
Noise level & $p$ & $K$ & $Q$ & $\hat{p}$  & $\hat{K}$\\
\hline
low     &0.45 $\pm$ 0.03    &22.84 $\pm$ 0.62    &  19.19 & 0.44& 22  \\ \hline
% medium  &0.20 $\pm$ 0.02    &25.90 $\pm$ 1.12  &0.23 $\pm$ 0.01    &21.72$\pm$ 0.03   & 10.25 & 0.27 & 25 \\ \hline
high    &0.13 $\pm$ 0.01    &17.73 $\pm$ 0.71    & 3.76 & 0.13 & 18 \\ \hline
\end{tabular}

% \begin{tabular}{|p{1.3cm}|| p{2cm}|p{2cm}|p{1.5cm}|p{1cm}|} \hline
%  \hline
% Spectrum &$p$                & $K$                 & $(Q_{K}, K_{\ell})$ & p(Q,K)\\
% \hline
% SPDC     &0.45 $\pm$ 0.03    &22.84 $\pm$ 0.62     &  (16.7 ,22 ) & 0.42 \\
% Normal   &0.42 $\pm$ 0.02    &20.00 $\pm$ 0.32     &  & \\
% \hline
% SPDC     &0.20 $\pm$ 0.02    &25.90 $\pm$ 1.12     & (10.25,25) & 0.27\\
% Normal   &0.23 $\pm$ 0.01    &21.72$\pm$ 0.03      &   &\\
%  \hline
% SPDC     &0.13 $\pm$ 0.01    &17.73 $\pm$ 0.71     &(3.45, 18) & 0.12\\
% Normal   &0.13 $\pm$ 0.01    &17.18  $\pm$ 0.34    & &\\
% \hline
% \end{tabular}
\caption{Measured purity ($p$) and dimensionality ($K$), under low and high noise levels, compared to estimates from other methods. Here $Q$ is the average quantum contrast.}
\label{tab:purity_dimen_summary}
\end{table}

\section{\textbf{Discussion and Conclusion}}

A measure of dimensionality and purity, particularly in the presence of (inevitable) deleterious noise that degrades the purity, is crucial for many quantum protocols and studies.  For example, there is a minimum purity needed to witness entanglement in a given dimension \cite{peres1996separability, horodecki1997separability, horodecki2009quantum}, setting the transition from separable to entangled states.  Likewise, knowing the purity is important in entanglement distillation processes since it informs whether the noise can be removed for a given dimension \cite{vollbrecht2003efficient, horodecki1999reduction}, while in entanglement based quantum communication there is a minimum purity \cite{Collins2002} associated with security \cite{ekert1991quantum}.  In turn, the dimensionality sets the information capacity of the state for quantum information processing and the error tolerance in quantum communication protocols, while high-dimensional states are important for fundamental tests of quantum mechanics where qubits will not suffice \cite{klyachko2008simple,lapkiewicz2011experimental}.  While a full quantum state tomography would also provide the necessary information, such a measurement would take several days for the states studied here, as compared to less than an hour with our technique.

%In most laboratory and all real-world environments, noise mechanisms that degrade the purity of entanglement are inevitable, and thus tools that inferred both dimensionality and purity are crucial for practical quantum systems.  Our technique solves this problem by using information extracted from various parts of the Hilbert space to determine the dimensions and purity of the state. Each measurement is reminiscent to a classifier that determines whether there is any entanglement within a given subspace, and to what degree, via a visibility that scales from 0 to 1. As such, the combination of  many such measurements is then used to quantify the dimensions of the system accurately.  

Unlike a conventional Schmidt decomposition, we do not assume the state is pure, and the dimension extracted from our technique is conditioned on the presence of entanglement: a maximally mixed and maximally entangled system cannot yield the same result.  While our approach would benefit from knowledge of the modal spectrum, which can be measured very quickly for OAM \cite{Pires2010,kulkarni2017single}, the outcome on purity and dimensionality are only modestly affected by typical spectrum shapes (see Supplementary Information), e.g., in our examples the uncertainty in dimensionality is $\approx 5\%$ with knowledge of the spectrum, increasing to $\approx 10\%$ without.  In this work we used OAM as our demonstration example, but reiterate that the projections are applicable to general states in the spatial mode basis. 

In summary, we have developed a simple yet powerful technique to measure the dimensionality and purity of high dimensional entangled photonic quantum systems. Our approach is robust, fast, and provides quantitative values rather than bounds or witnesses, and works on both pure and mixed states. Our scheme exploits visibility in fringes after joint projections, making it fast and easy to implement, returning the key parameters of the system in a fraction of the time that a full quantum state tomography would take. We believe that this tool will be useful to the active research in high-dimensional spatial mode entanglement and foster its wide-spread deployment in quantum based protocols.
\vspace{0.2 cm}

%%%%%%%%%%%%%%%%%%%%%%%%%%%%%%%%%%%%%%%%%%%%%%%%%%
%%% Additional info
%%%%%%%%%%%%%%%%%%%%%%%%%%%%%%%%%%%%%%%%%%%%%%%%%%

%%%%%%
\section*{\textbf{Acknowledgements}}
The authors express their gratitude to Bienvenue Ndagano for his inputs. I.N. would like to acknowledge the Department of Science and Technology (South Africa) for funding.
\vspace{0.2 cm}

%%%%%
\section*{\textbf{Author contributions}}
The experiment was performed by I.N. and V.R.F., the theory was developed by I.N., F.Z,. H.H., and J.L., the data analysis was performed by I.N., V.R.F. and A.F. and the experiment was conceived by I.N., V.R.F., and A.F. All authors contributed to the writing of the manuscript. A.F. supervised the project.

%%%%%
\section*{\textbf{ Competing Interests statement}}
The authors declare not competing interests.

%%%%%%%%%%%%%%%%%%%%%%%%%%%%%%%%%%%%%%%%%%%%%%%%%%
\section*{Methods}\label{sec:holo_gen}
%%%%%%%%%%%%%%%%%%%%%%%%%%%%%%%%%%%%%%%%%%%%%%%%%%
%%%%%%%%%%%%%%%%%%%%%%%%%%%%%%%%%%%%%%%%%%%%%%%%%%

\noindent\textbf{High dimensionsal state projections.}  Our analysers project onto the high dimensional Hilbert space, $\mathcal{H}_d$, mapping onto the states in Eq. (\ref{eq:NFracState2}), i.e $\ket{M, \alpha}_n$, repeated here as 
\begin{equation}
\ket{M,\alpha}_n = \mathcal{N} \sum^{d-1}_{j=0} c_{w_j,M}^n(\alpha) \ket{j},
\label{eq:NFracState3}
\end{equation}
\noindent composed from coherent superpositions of basis states $\ket{j}\in\{\ket{j}, j = 0,1..d-1 \}$ with tune-able phases and amplitudes
\begin{equation}\label{eq:SPPnAmpltudes}
%c_{\ell,M}^n(\alpha) =  \mathcal{N} \exp (-i \pi \ell(n-1)/n)  A_{\ell}^n   c_{\ell, M}(\alpha),
c_{w_j,M}^n(\alpha) =  \mathcal{N} e^{-i \pi w_j(n-1)/n}  A_{w_j}^n   c_{w_j, M}(\alpha),
\end{equation} 
and where $w_j=j-(d-1)/2$ and the factors
%%% equation
\begin{equation}
c_{w_j, M}(\alpha) = -\frac{i e^{-i w_j \alpha}  }{ \pi  (M-w_j)}.
\label{eq:fracCoeff}
\end{equation}
%%% 
and
%%% equation
\begin{align}
A_{w_j}^n  =& \left\{\begin{array}{cc}
1, & \text{mod}\left\{ w_j, n \right\} = 0 \\[1mm]
0, & \text{otherwise} \\
\end{array} \right.
\label{eq:Selec}.
\end{align}
%%%
Here, $c_{w_k, M}(\alpha)$ controls the relative phases and amplitudes of the  eigenmodes and $A_{w_j}^n$ modulates the coefficients' amplitudes while $\alpha\in[0, 2\pi/n]$. The spectrum given by Eq. \ref{eq:SPPnAmpltudes} can be tuned by carefully selecting $n$, therefore enabling precise control of the subspaces that will be probed.

In the OAM basis, i.e $\ket{\ell} \in \mathcal{H}_d$, the index $w_j$ can be replaced with the index $\ell\in \mathcal{Z}$. The mode projectors can be constructed from spiral phase profiles having the transmission function
%%% equation
\begin{equation}\label{eq:NFracState2}
U_{n}(\phi, \alpha) = \mathcal{M}\sum^{n-1}_{k=0} \exp \left( i \Phi_M\left( \phi;\beta{_k}\oplus\alpha \right) \right),
\end{equation}
%%%
\noindent that is constructed from superpositions of fractional OAM modes (See Supplementary Information) \cite{gotte2007quantum, oemrawsingh2005experimental},
%%%equation
\begin{equation}\label{eq:phaseSPP}
\exp \left(i \Phi_{M}(\phi; \alpha) \right) = 
\begin{cases}
e^{i M \left( 2\pi +\phi - \alpha \right)} & \quad 0 \leq \phi < \alpha \\
e^{i M \left( \phi - \alpha \right)} & \quad \alpha \leq \phi < 2\pi
\end{cases},
\end{equation}
%%%
having the same charge, $M$, but rotated by an angle $\beta_k\oplus\alpha=\text{mod}\left\{ \beta_k + \alpha, 2\pi \right \}$ for $\beta_k=\frac{2\pi}{n}k$, as illustrated in Figs.~\ref{fig:ConceptFig} (b) and (c) for $n=3$ and $n=7$, respectively. Here, $\phi$ is the azimuthal coordinate and $\mathcal{M}$ a normalization constant.

\noindent\textbf{Experimental setup.} The experimental setup for the generation and measurement of entangled photons is illustrated schematically in Fig.~\ref{fig:ConceptFig} (d). A potassium-titanium-phosphate (PPKTP) type I nonlinear crystal (NLC) was pumped with a 405 nm wavelength diode laser. The crystal temperature was set to obtain co-linear signal and idler entangled SPDC photons centred at a wavelength 810 nm. The photon pairs were then separated in path using a 50:50 beam splitter (BS). Each entangled photon was imaged onto a spatial light modulator (SLM) using a $4f$ telescope ($f_1$ and $f_2$ having focal lengths of 100 mm and  500 mm, respectively), then subsequently coupled into single mode fibers with a second $4f$ telescope (lenses $f_3$ and $f_4$ having focal lengths of 750 mm and 2 mm, respectively) and finally detected with avalanche photo-detectors (APDs). Signals from each arm were measured in coincidences within a 25 ns coincidence window. The entangled photons were filtered with 10 nm bandpass filters centered at a wavelength of 810 nm. For our experimental demonstration, we restrict our measurements to a specific optical setup and we varied the purity of the entangled state by introducing background noise in the form of white light. To obtain a high purity state (p=0.45 in K=22 dimensions), we had to reduce the laser power using an ND filter such as to reduce multi-photon emission events, known to have an impact on the purity of the SPDC photons \cite{zhu2019high}. 

\noindent\textbf{Holographic fractional OAM mode projections.}  The conventional method for generating fractional OAM modes involves imprinting an azimuthally dependent phase retardation onto an incoming field with spiral phase plates \cite{oemrawsingh2004half} or digital holograms \cite{leach2004observation}. Here, we employed dynamic phase control of liquid crystal spatial light modulators (SLMs) \cite{rosales2017shape} to generate rotated fractional OAM phase masks to modulate the transverse plane of photons. We achieved this by encoding grey-scale holograms with the phase profile of a transmission function corresponding to the desired projection mode. Accordingly, we prepared analysers with superpositions of multiple rotated fractional spiral phases having a transmission function given by eq. \ref{eq:NFracState2}, with the azimuthal coordinate $\phi=\tan^{-1}(\frac{y}{x})$ and $(x,y)$ the coordinates of each pixel. We generated holograms with the desired phase and subsequently added them to a blazed grating for a final hologram mapped as
%%% equation
\begin{align}\label{eqn:Finalholo}
\Phi_{\text{SLM}}(x,y)=\text{mod}\left\{\text{arg}\left( U_n(\alpha, \phi) \right) \right.\nonumber\\ 
\left.+ G_x x +  G_y y, 2\pi \right\},
\end{align}
%%%
\noindent where $G_{x,y}$ are the grating wavenumbers in the $x$ and $y$ directions, respectively. An example of the phase hologram generated of a superposition of three fractional OAM spiral phases ($n=3$ with $M=1.5$), and its corresponding rotated complex-conjugate are shown as inserts in Fig.~\ref{fig:ConceptFig} (d).

\noindent\textbf{Optimal purity and dimensionality calculation.}  Using the fact that the visibility obtained for each analyser is affected by the dimensionality and purity of the input state, we describe the procedure for determining their values for a given entangled quantum system, assuming it can be modelled by the isotropic state in Eq.~(\ref{eq:isostate}). We measure the probability curves for N analysers each with $n=1,3,...,2N-1$, and compute their corresponding visibilities $V_n :=V_n(p,K)$. This results in a set of $N$ nonlinear equations that depend on the parameters $p$ and $K$. We then determine the optimal $(p,K)$ pair that best fit the function $V_n(p,K)$ to all $N$ measured visibilies by employing the method of least squares (LSF), which aims to minimise the objective function
%%% Eq
\begin{equation}
\chi^2(p,K)=\sum_{i=1}^{N} | V_{2i-1}^\text{The.}(p, K)-V^\text{Exp.}_{2i-1}|^2.
\end{equation}
%%%%%
\noindent where are the terms in the summation are the residuals (absolute errors) for each $n=2i-1$ visibility measurement with respect to the theory. 

% %%% fig
% \begin{figure}[t]
% \centering
% \includegraphics[width=1\linewidth]{Setup_sup.jpg}
% \caption{\IN{\textbf{Experimental setup.} (a)-(d) Phase profiles of  fractional OAM analysers for $n=1,2,3$, and $4$, all with charges $M=n/2$, respectively. (e) Experimental setup. SLM: spatial light modulator; L: lens; NC: nonlinear crystal; BS: beam splitter; M: mirror; SMF: single mode fibre; APD: avalanche photodiode; BPF: bandpass filter; s: signal; i: idler; C.C: coincidence counter. Inset: illustration of the hologram generation.}}\label{fig:setup}
% \end{figure}
% %%%

%%%%%%%%%%%%%%%%%%%%%%%%%%%%%%%%%%%%%%%%%%%%%%%%%%
%%% References
%%%%%%%%%%%%%%%%%%%%%%%%%%%%%%%%%%%%%%%%%%%%%%%%%%

\bibliography{mybibfile}

\begin{thebibliography}{10}

\bibitem{cozzolino2019high}
D.~Cozzolino, B.~Da~Lio, D.~Bacco, and L.~K. Oxenl{\o}we, ``High-dimensional
  quantum communication: Benefits, progress, and future challenges,'' {\em
  Advanced Quantum Technologies}, vol.~2, no.~12, p.~1900038, 2019.

\bibitem{erhard2017quantum}
M.~Erhard, M.~Malik, and A.~Zeilinger, ``A quantum router for high-dimensional
  entanglement,'' {\em Quantum Science and Technology}, vol.~2, no.~1,
  p.~014001, 2017.

\bibitem{babazadeh2017high}
A.~Babazadeh, M.~Erhard, F.~Wang, M.~Malik, R.~Nouroozi, M.~Krenn, and
  A.~Zeilinger, ``High-dimensional single-photon quantum gates: concepts and
  experiments,'' {\em Physical review letters}, vol.~119, no.~18, p.~180510,
  2017.

\bibitem{deng2017quantum}
F.-G. Deng, B.-C. Ren, and X.-H. Li, ``Quantum hyperentanglement and its
  applications in quantum information processing,'' {\em Science bulletin},
  vol.~62, no.~1, pp.~46--68, 2017.

\bibitem{gisin1997optimal}
N.~Gisin and S.~Massar, ``Optimal quantum cloning machines,'' {\em Physical
  review letters}, vol.~79, no.~11, p.~2153, 1997.

\bibitem{bouchard2017high}
F.~Bouchard, R.~Fickler, R.~W. Boyd, and E.~Karimi, ``High-dimensional quantum
  cloning and applications to quantum hacking,'' {\em Science advances},
  vol.~3, no.~2, p.~e1601915, 2017.

\bibitem{ecker2019overcoming}
S.~Ecker, F.~Bouchard, L.~Bulla, F.~Brandt, O.~Kohout, F.~Steinlechner,
  R.~Fickler, M.~Malik, Y.~Guryanova, R.~Ursin, {\em et~al.}, ``Overcoming
  noise in entanglement distribution,'' {\em Physical Review X}, vol.~9, no.~4,
  p.~041042, 2019.

\bibitem{dixon2008gigahertz}
A.~Dixon, Z.~Yuan, J.~Dynes, A.~Sharpe, and A.~Shields, ``Gigahertz decoy
  quantum key distribution with 1 mbit/s secure key rate,'' {\em Optics
  express}, vol.~16, no.~23, pp.~18790--18797, 2008.

\bibitem{barreiro2008beating}
J.~T. Barreiro, T.-C. Wei, and P.~G. Kwiat, ``Beating the channel capacity
  limit for linear photonic superdense coding,'' {\em Nature physics}, vol.~4,
  no.~4, p.~282, 2008.

\bibitem{Agnew2011}
M.~Agnew, J.~Leach, M.~McLaren, F.~S. Roux, and R.~W. Boyd, ``{Tomography of
  the quantum state of photons entangled in high dimensions},'' {\em Physical
  Review A}, vol.~84, p.~062101, 2011.

\bibitem{giovannini2013characterization}
D.~Giovannini, J.~Romero, J.~Leach, A.~Dudley, A.~Forbes, and M.~J. Padgett,
  ``Characterization of high-dimensional entangled systems via mutually
  unbiased measurements,'' {\em Physical Review Letters}, vol.~110, no.~14,
  p.~143601, 2013.

\bibitem{bavaresco2018measurements}
J.~Bavaresco, N.~H. Valencia, C.~Kl{\"o}ckl, M.~Pivoluska, P.~Erker, N.~Friis,
  M.~Malik, and M.~Huber, ``Measurements in two bases are sufficient for
  certifying high-dimensional entanglement,'' {\em Nature Physics},
  pp.~1745--2481, 2018.

\bibitem{vaziri2002experimental}
A.~Vaziri, G.~Weihs, and A.~Zeilinger, ``Experimental two-photon,
  three-dimensional entanglement for quantum communication,'' {\em Physical
  Review Letters}, vol.~89, no.~24, p.~240401, 2002.

\bibitem{groblacher2006experimental}
S.~Gr{\"o}blacher, T.~Jennewein, A.~Vaziri, G.~Weihs, and A.~Zeilinger,
  ``Experimental quantum cryptography with qutrits,'' {\em New Journal of
  Physics}, vol.~8, no.~5, p.~75, 2006.

\bibitem{dada2011experimental}
A.~C. Dada, J.~Leach, G.~S. Buller, M.~J. Padgett, and E.~Andersson,
  ``Experimental high-dimensional two-photon entanglement and violations of
  generalized bell inequalities,'' {\em Nature Physics}, vol.~7, no.~9, p.~677,
  2011.

\bibitem{oemrawsingh2005experimental}
S.~Oemrawsingh, X.~Ma, D.~Voigt, A.~Aiello, E.~t. Eliel, J.~Woerdman, {\em
  et~al.}, ``Experimental demonstration of fractional orbital angular momentum
  entanglement of two photons,'' {\em Physical review letters}, vol.~95,
  no.~24, p.~240501, 2005.

\bibitem{oemrawsingh2006high}
S.~Oemrawsingh, J.~de~Jong, X.~Ma, A.~Aiello, E.~Eliel, J.~Woerdman, {\em
  et~al.}, ``High-dimensional mode analyzers for spatial quantum
  entanglement,'' {\em Physical Review A}, vol.~73, no.~3, p.~032339, 2006.

\bibitem{gotte2007quantum}
J.~B. G{\"o}tte, S.~Franke-Arnold, R.~Zambrini, and S.~M. Barnett, ``Quantum
  formulation of fractional orbital angular momentum,'' {\em Journal of Modern
  Optics}, vol.~54, no.~12, pp.~1723--1738, 2007.

\bibitem{huang2018various}
H.-C. Huang, ``Various angle periods of parabolic coincidence fringes in
  violation of the bell inequality with high-dimensional two-photon
  entanglement,'' {\em Phys. Rev. A}, vol.~98, p.~053856, Nov 2018.

\bibitem{dada2011bell}
A.~C. Dada and E.~Andersson, ``On bell inequality violations with
  high-dimensional systems,'' {\em International Journal of Quantum
  Information}, vol.~9, no.~07n08, pp.~1807--1823, 2011.

\bibitem{romero2013tailored}
J.~Romero, D.~Giovannini, D.~Tasca, S.~Barnett, and M.~Padgett, ``Tailored
  two-photon correlation and fair-sampling: a cautionary tale,'' {\em New
  Journal of Physics}, vol.~15, no.~8, p.~083047, 2013.

\bibitem{pors2008shannon}
J.~Pors, S.~Oemrawsingh, A.~Aiello, M.~Van~Exter, E.~Eliel, J.~Woerdman, {\em
  et~al.}, ``Shannon dimensionality of quantum channels and its application to
  photon entanglement,'' {\em Physical review letters}, vol.~101, no.~12,
  p.~120502, 2008.

\bibitem{zhu2019high}
F.~Zhu, M.~Tyler, N.~H. Valencia, M.~Malik, and J.~Leach, ``Are
  high-dimensional entangled states robust to noise?,'' {\em arXiv preprint
  arXiv:1908.08943}, 2019.

\bibitem{fickler2012quantum}
R.~Fickler, R.~Lapkiewicz, W.~N. Plick, M.~Krenn, C.~Schaeff, S.~Ramelow, and
  A.~Zeilinger, ``Quantum entanglement of high angular momenta,'' {\em
  Science}, vol.~338, no.~6107, pp.~640--643, 2012.

\bibitem{krenn2015twisted}
M.~Krenn, J.~Handsteiner, M.~Fink, R.~Fickler, and A.~Zeilinger, ``Twisted
  photon entanglement through turbulent air across vienna,'' {\em Proceedings
  of the National Academy of Sciences}, vol.~112, no.~46, pp.~14197--14201,
  2015.

\bibitem{erhard2018twisted}
M.~Erhard, R.~Fickler, M.~Krenn, and A.~Zeilinger, ``Twisted photons: new
  quantum perspectives in high dimensions,'' {\em Light: Science \&
  Applications}, vol.~7, no.~3, pp.~17146--17146, 2018.

\bibitem{fickler2014interface}
R.~Fickler, R.~Lapkiewicz, M.~Huber, M.~P. Lavery, M.~J. Padgett, and
  A.~Zeilinger, ``Interface between path and orbital angular momentum
  entanglement for high-dimensional photonic quantum information,'' {\em Nature
  communications}, vol.~5, no.~1, pp.~1--6, 2014.

\bibitem{forbes2019quantum}
A.~Forbes and I.~Nape, ``Quantum mechanics with patterns of light: Progress in
  high dimensional and multidimensional entanglement with structured light,''
  {\em AVS Quantum Science}, vol.~1, no.~1, p.~011701, 2019.

\bibitem{molina2007twisted}
G.~Molina-Terriza, J.~P. Torres, and L.~Torner, ``Twisted photons,'' {\em
  Nature Physics}, vol.~3, no.~5, pp.~305--310, 2007.

\bibitem{wang2005quantum}
C.~Wang, F.-G. Deng, Y.-S. Li, X.-S. Liu, and G.~L. Long, ``Quantum secure
  direct communication with high-dimension quantum superdense coding,'' {\em
  Physical Review A}, vol.~71, no.~4, p.~044305, 2005.

\bibitem{Malik2016}
M.~Malik, M.~Erhard, M.~Huber, M.~Krenn, R.~Fickler, and A.~Zeilinger,
  ``{Multi-photon entanglement in high dimensions},'' {\em Nature Photonics},
  vol.~10, pp.~248--252, 2016.

\bibitem{goyal2014qudit}
S.~K. Goyal, P.~E. Boukama-Dzoussi, S.~Ghosh, F.~S. Roux, and T.~Konrad,
  ``Qudit-teleportation for photons with linear optics,'' {\em Scientific
  reports}, vol.~4, p.~4543, 2014.

\bibitem{zhang2017simultaneous}
Y.~Zhang, M.~Agnew, T.~Roger, F.~S. Roux, T.~Konrad, D.~Faccio, J.~Leach, and
  A.~Forbes, ``Simultaneous entanglement swapping of multiple orbital angular
  momentum states of light,'' {\em Nature Communications}, vol.~8, no.~1,
  p.~632, 2017.

\bibitem{bornman2019ghost}
N.~Bornman, M.~Agnew, F.~Zhu, A.~Vall{\'e}s, A.~Forbes, and J.~Leach, ``Ghost
  imaging using entanglement-swapped photons,'' {\em npj Quantum Information},
  vol.~5, no.~1, pp.~1--6, 2019.

\bibitem{jack2009holographic}
B.~Jack, J.~Leach, J.~Romero, S.~Franke-Arnold, M.~Ritsch-Marte, S.~M. Barnett,
  and M.~J. Padgett, ``Holographic ghost imaging and the violation of a bell
  inequality,'' {\em Phys. Rev. Lett.}, vol.~103, p.~083602, 2009.

\bibitem{chen2014quantum}
L.~Chen, J.~Lei, and J.~Romero, ``Quantum digital spiral imaging,'' {\em Light:
  Science \& Applications}, vol.~3, no.~3, pp.~e153--e153, 2014.

\bibitem{bornman2019ghostaa}
N.~Bornman, S.~Prabhakar, A.~Vall{\'e}s, J.~Leach, and A.~Forbes, ``Ghost
  imaging with engineered quantum states by hong--ou--mandel interference,''
  {\em New Journal of Physics}, vol.~21, no.~7, p.~073044, 2019.

\bibitem{mafu2013higher}
M.~Mafu, A.~Dudley, S.~Goyal, D.~Giovannini, M.~McLaren, M.~J. Padgett,
  T.~Konrad, F.~Petruccione, N.~L{\"u}tkenhaus, and A.~Forbes,
  ``Higher-dimensional orbital-angular-momentum-based quantum key distribution
  with mutually unbiased bases,'' {\em Physical Review A}, vol.~88, no.~3,
  p.~032305, 2013.

\bibitem{steinlechner2017distribution}
F.~Steinlechner, S.~Ecker, M.~Fink, B.~Liu, J.~Bavaresco, M.~Huber, T.~Scheidl,
  and R.~Ursin, ``Distribution of high-dimensional entanglement via an
  intra-city free-space link,'' {\em Nature communications}, vol.~8, p.~15971,
  2017.

\bibitem{liu2020multidimensional}
J.~Liu, I.~Nape, Q.~Wang, A.~Vall{\'e}s, J.~Wang, and A.~Forbes,
  ``Multidimensional entanglement transport through single-mode fiber,'' {\em
  Science advances}, vol.~6, no.~4, p.~eaay0837, 2020.

\bibitem{cao2020distribution}
H.~Cao, S.-C. Gao, C.~Zhang, J.~Wang, D.-Y. He, B.-H. Liu, Z.-W. Zhou, Y.-J.
  Chen, Z.-H. Li, S.-Y. Yu, {\em et~al.}, ``Distribution of high-dimensional
  orbital angular momentum entanglement over a 1 km few-mode fiber,'' {\em
  Optica}, vol.~7, no.~3, pp.~232--237, 2020.

\bibitem{cozzolino2019air}
D.~Cozzolino, E.~Polino, M.~Valeri, G.~Carvacho, D.~Bacco, N.~Spagnolo, L.~K.
  Oxenl{\o}we, and F.~Sciarrino, ``Air-core fiber distribution of hybrid vector
  vortex-polarization entangled states,'' {\em Advanced Photonics}, vol.~1,
  no.~4, p.~046005, 2019.

\bibitem{cozzolino2019orbital}
D.~Cozzolino, D.~Bacco, B.~Da~Lio, K.~Ingerslev, Y.~Ding, K.~Dalgaard,
  P.~Kristensen, M.~Galili, K.~Rottwitt, S.~Ramachandran, {\em et~al.},
  ``Orbital angular momentum states enabling fiber-based high-dimensional
  quantum communication,'' {\em Physical Review Applied}, vol.~11, no.~6,
  p.~064058, 2019.

\bibitem{leach2012secure}
J.~Leach, E.~Bolduc, D.~J. Gauthier, and R.~W. Boyd, ``Secure information
  capacity of photons entangled in many dimensions,'' {\em Physical Review A},
  vol.~85, no.~6, p.~060304, 2012.

\bibitem{zhang2016engineering}
Y.~Zhang, F.~S. Roux, T.~Konrad, M.~Agnew, J.~Leach, and A.~Forbes,
  ``Engineering two-photon high-dimensional states through quantum
  interference,'' {\em Science Advances}, vol.~2, no.~2, p.~e1501165, 2016.

\bibitem{torres2003quantum}
J.~Torres, A.~Alexandrescu, and L.~Torner, ``Quantum spiral bandwidth of
  entangled two-photon states,'' {\em Physical Review A}, vol.~68, no.~5,
  p.~050301, 2003.

\bibitem{miatto2011full}
F.~M. Miatto, A.~M. Yao, and S.~M. Barnett, ``Full characterization of the
  quantum spiral bandwidth of entangled biphotons,'' {\em Physical Review A},
  vol.~83, no.~3, p.~033816, 2011.

\bibitem{Romero2012}
J.~Romero, D.~Giovannini, S.~Franke-Arnold, S.~M. Barnett, and M.~J. Padgett,
  ``Increasing the dimension in high-dimensional two-photon orbital angular
  momentum entanglement,'' {\em Phys. Rev. A}, vol.~86, p.~012334, Jul 2012.

\bibitem{Terriza}
G.~Molina-Terriza, J.~P. Torres, and L.~Torner, ``Management of the angular
  momentum of light: Preparation of photons in multidimensional vector states
  of angular momentum,'' {\em Phys. Rev. Lett.}, vol.~88, p.~013601, Dec 2001.

\bibitem{zhang2014simulating}
Y.~Zhang, M.~Mclaren, F.~S. Roux, and A.~Forbes, ``Simulating quantum state
  engineering in spontaneous parametric down-conversion using classical
  light,'' {\em Optics Express}, vol.~22, no.~14, pp.~17039--17049, 2014.

\bibitem{krenn2014generation}
M.~Krenn, M.~Huber, R.~Fickler, R.~Lapkiewicz, S.~Ramelow, and A.~Zeilinger,
  ``Generation and confirmation of a (100$\times$ 100)-dimensional entangled
  quantum system,'' {\em Proceedings of the National Academy of Sciences},
  vol.~111, no.~17, pp.~6243--6247, 2014.

\bibitem{peres1996separability}
A.~Peres, ``Separability criterion for density matrices,'' {\em Physical Review
  Letters}, vol.~77, no.~8, p.~1413, 1996.

\bibitem{horodecki1997separability}
P.~Horodecki, ``Separability criterion and inseparable mixed states with
  positive partial transposition,'' {\em Physics Letters A}, vol.~232, no.~5,
  pp.~333--339, 1997.

\bibitem{horodecki2009quantum}
R.~Horodecki, P.~Horodecki, M.~Horodecki, and K.~Horodecki, ``Quantum
  entanglement,'' {\em Reviews of modern physics}, vol.~81, no.~2, p.~865,
  2009.

\bibitem{vollbrecht2003efficient}
K.~G.~H. Vollbrecht and M.~M. Wolf, ``Efficient distillation beyond qubits,''
  {\em Physical Review A}, vol.~67, no.~1, p.~012303, 2003.

\bibitem{horodecki1999reduction}
M.~Horodecki and P.~Horodecki, ``Reduction criterion of separability and limits
  for a class of distillation protocols,'' {\em Physical Review A}, vol.~59,
  no.~6, p.~4206, 1999.

\bibitem{Collins2002}
D.~Collins, N.~Gisin, N.~Linden, S.~Massar, and S.~Popescu, ``Bell inequalities
  for arbitrarily high-dimensional systems,'' {\em Phys. Rev. Lett.}, vol.~88,
  p.~040404, Jan 2002.

\bibitem{ekert1991quantum}
A.~K. Ekert, ``Quantum cryptography based on bell’s theorem,'' {\em Physical
  review letters}, vol.~67, no.~6, p.~661, 1991.

\bibitem{klyachko2008simple}
A.~A. Klyachko, M.~A. Can, S.~Binicio{\u{g}}lu, and A.~S. Shumovsky, ``Simple
  test for hidden variables in spin-1 systems,'' {\em Physical review letters},
  vol.~101, no.~2, p.~020403, 2008.

\bibitem{lapkiewicz2011experimental}
R.~Lapkiewicz, P.~Li, C.~Schaeff, N.~K. Langford, S.~Ramelow, M.~Wie{\'s}niak,
  and A.~Zeilinger, ``Experimental non-classicality of an indivisible quantum
  system,'' {\em Nature}, vol.~474, no.~7352, pp.~490--493, 2011.

\bibitem{Pires2010}
H.~D.~L. Pires, H.~C.~B. Florijn, and M.~P. van Exter, ``Measurement of the
  spiral spectrum of entangled two-photon states,'' {\em Phys. Rev. Lett.},
  vol.~104, p.~020505, 2010.

\bibitem{kulkarni2017single}
G.~Kulkarni, R.~Sahu, O.~S. Maga{\~n}a-Loaiza, R.~W. Boyd, and A.~K. Jha,
  ``Single-shot measurement of the orbital-angular-momentum spectrum of
  light,'' {\em Nature communications}, vol.~8, no.~1, pp.~1--8, 2017.

\bibitem{oemrawsingh2004half}
S.~Oemrawsingh, E.~Eliel, J.~Woerdman, E.~Verstegen, J.~Kloosterboer, {\em
  et~al.}, ``Half-integral spiral phase plates for optical wavelengths,'' {\em
  Journal of Optics A: Pure and Applied Optics}, vol.~6, no.~5, p.~S288, 2004.

\bibitem{leach2004observation}
J.~Leach, E.~Yao, and M.~J. Padgett, ``Observation of the vortex structure of a
  non-integer vortex beam,'' {\em New Journal of Physics}, vol.~6, no.~1,
  p.~71, 2004.

\bibitem{rosales2017shape}
C.~Rosales-Guzm{\'a}n and A.~Forbes, {\em How to shape light with spatial light
  modulators}.
\newblock SPIE Press, 2017.

\end{thebibliography}
\bibliographystyle{ieeetr}

\section{Schmidt number}
The Schmidt number ($K$) quantifies the minimum number of basis states taken from an overcomplete set, required to fully describe a quantum system. For example, in the OAM basis, the product state, $\ket{\ell} \ket{-\ell}$, for some integer $\ell$, can be used to represent a system of two photons,
%%% eq
\begin{equation}
	\ket{\Psi}=\sum_{\ell=-L}^L \lambda_{\ell}\ket{\ell} \ket{-\ell},
	\label{eq: spdc}
\end{equation}
%%%
\noindent where $|\lambda_{\ell}|^2$ is the probability of detecting the biphoton state $\ket{\ell} \ket{-\ell}$. The Schmidt number of such a state can be obtained from
%%% eq
\begin{equation}
	K=\frac{\left( \sum_\ell |\lambda_\ell|^2 \right)^2}{\sum_\ell |\lambda_\ell|^4}.
	\label{eq: scmidt}
\end{equation}
%%%%
\noindent Another convenient measure is the width,  $\Delta \ell$, of the distribution  as two times the square root of its variance (second central moment), given by
\begin{equation}
	\Delta\ell=2\sqrt{ \frac{\sum_\ell |\ell|^2 \ |\lambda_\ell|^2 }{\sum_\ell |\lambda_\ell|^2} }.
\end{equation}

%%% figure
\begin{figure}[h]
	\centering
	\includegraphics[width=1\linewidth]{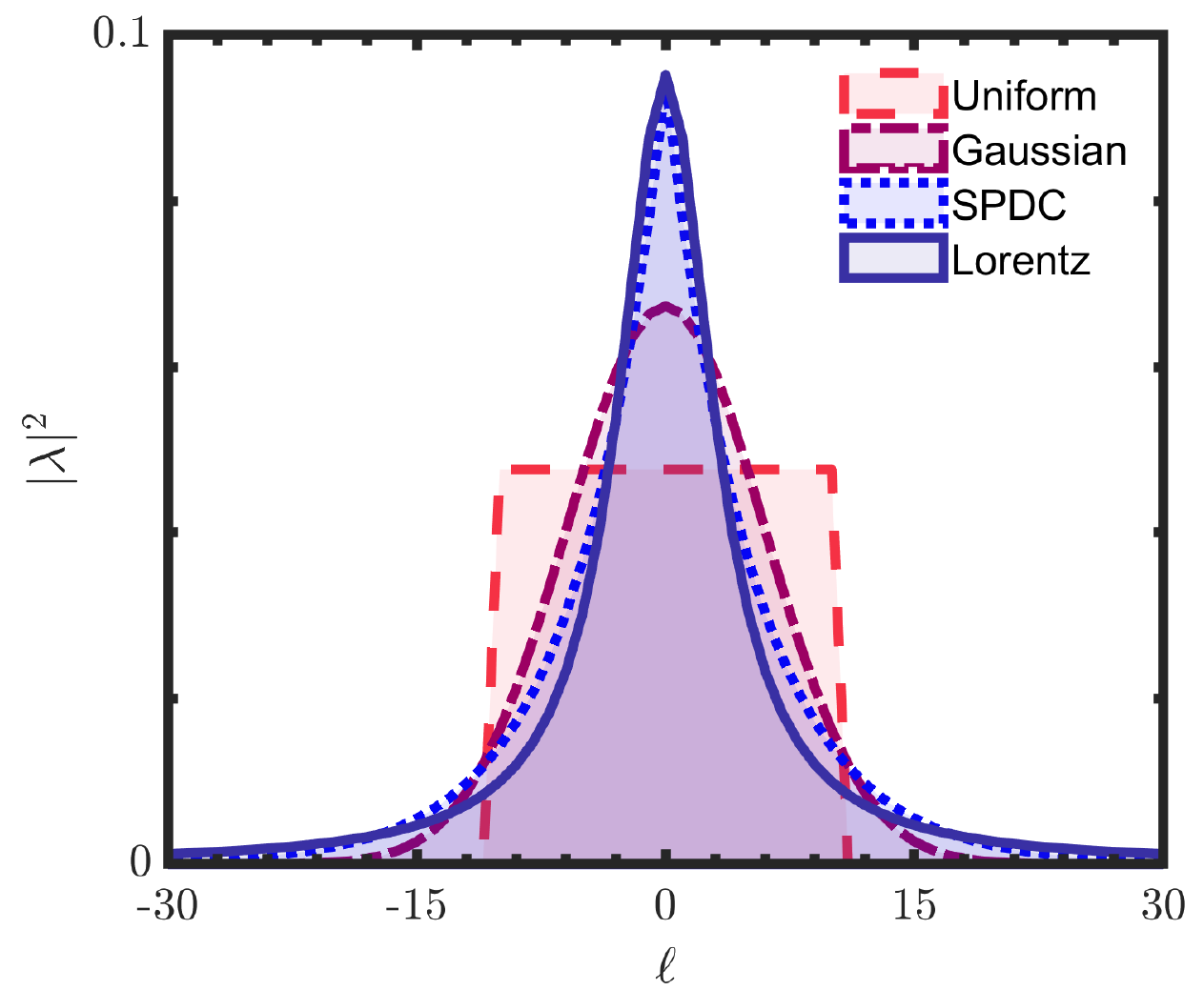}
	\caption{Examples of various OAM ($\ell$) distributions for a quantum source possessing OAM entanglement.}\label{fig:Spectrum}
\end{figure}
Both $K$ and $\Delta \ell$ depend on the shape of the distribution $|\lambda_\ell|^2$. 
Examples of various types of distributions for $|\lambda_\ell|^2$ are shown in Fig. \ref{fig:Spectrum} for $K=21$. The distributions are: a square distribution, corresponding to a uniform (maximally entangled state) within a given $\ell$-range,
%%% eq
\begin{equation}
	\lambda_{\ell}=1/ \sqrt{2L +1}, \ |\ell| \leq L;
\end{equation}
%%%
a Gaussian (normal) distribution
%%% eq
\begin{equation}
	|\lambda_{\ell_g}|^2 \propto \exp\left[-\frac{|\ell|^2}{\gamma_G^2}\right],
\end{equation}
%%%
where $\gamma_G$ scales with the width of the distribution; a SPDC source \cite{miatto2011full, torres2003quantum}
%%% eq
\begin{equation}
	|\lambda_{\ell_S}|^2 \propto \left(\frac{2\gamma_{S}^2}{1+\gamma_{S}^2}\right)^{|\ell|},
\end{equation}
%%%
where  $\gamma_{S}$ is determined by the experimental conditions; and a Lorentz distribution
%%% eq
\begin{equation}
	|\lambda_{\ell_L}|^2 \propto  \frac{1}{\pi \gamma_L\left(1+\frac{\ell^2}{\gamma_L^2}\right)},
\end{equation}
%%%
where $\gamma_L$ is a scaling parameter. 

Note that all distributions in Fig. \ref{fig:Spectrum} have the same $K$ value, but differ in $\Delta\ell$, with $\Delta\ell=12,11.8,14.6 \text{ and }41.45$ for the square, Gaussian, SPDC and Lorentz distributions, respectively.

In this work, we make use the SPDC and normal distributions. For convenience, we relate the the Schmidt number to the scaling parameters as, $\gamma_S \approx \sqrt{(K-1)/4}$ and $\gamma_G \approx 2.5066 K$ for the SPDC and normal distributions, respectively.

%%%%%%%%%%%%%%%%%%%%%%%%%%%%%%%%%%%%%%%%%%%%%%%%%%
%%%%%%%%%%%%%%%%%%%%%%%%%%%%%%%%%%%%%%%%%%%%%%%%%%
\section{Single fractional OAM analyser}\label{sec:Frac_OAM}
A single fractional OAM mode analyser can be represented on the high dimensional Hilbert space using the OAM basis modes  $\ket{\ell}\in\mathcal{H}_\infty$ as
%%% equation
\begin{equation}
	\ket{M, \alpha} = \sum_{\ell=-\infty}^{\infty} c_{\ell,M}(\alpha) \ket{\ell},
	\label{eq : fracstate}
\end{equation}
%%% equation
\noindent where the complex coefficients, $c_{\ell, M}(\alpha)$, are computed from the overlap integral, $\int e^{-i \ell \phi} e^{i\Phi_{M}(\phi; \alpha)} \ d\phi $. Here  $e^{i\Phi_{M}(\phi; \alpha)}$ is the azimuthally dependent mode characterizing the analyser orientated at an angle $\alpha$. Note that a complete decomposition would require an expansion onto a complete basis that  includes the radial component. For brevity, we restrict ourselves to the azimuthal degree of freedom, consistent with \cite{gotte2007quantum}.

By computing the overlap integral, one arrives at complex amplitudes
%%% equation
\begin{equation}
	c_{\ell,M}(\alpha) = -\frac{i e^{-i \ell \alpha  } \sin{\left( \mu\pi \right)} }{ \pi  (M-\ell)},
	\label{eq:fracCoeff}
\end{equation}
%%%
\noindent with $\mu$ representing the fractional part of the total charge $M$.  The detection probability for each OAM mode with charge $\ell$ is therefore 
%%% equation
\begin{equation}\label{SPPOAMDst}
	P_\ell = |c_{\ell,M}(\alpha)|^2 = \frac{\sin^2(\mu\pi)}{ \pi^2 (M-\ell)^2 },
\end{equation}
%%%
consistent with probability amplitudes computed in \cite{gotte2007quantum} for fractional OAM states.

%%%%%%%%%%%%%%%%%%%%%%%%%%%%%%%%%%%%%%%%%%%%%%%%%%
%%%%%%%%%%%%%%%%%%%%%%%%%%%%%%%%%%%%%%%%%%%%%%%%%%
\section{Superpositions of fractional OAM analysers}
We have shown that fractional OAM modes project onto the high dimensional state space of OAM modes with complex amplitudes given by Eq.~(\ref{eq:fracCoeff}). Next, we tailor new amplitudes and phases by superimposing rotated fractional OAM modes
%%% equation
\begin{equation}
	\ket{M,\alpha}_n = \mathcal{N}\sum^{n-1}_{k=0}\ket{M,\beta{_k}\oplus\alpha},
	\label{eq:NFracStateSup2}
\end{equation}
%%%%
\noindent where $\mathcal{N}$ is a normalization constant. Each fractional mode in this superposition has the same charge, $M$, but is rotated by an angle $\beta_k\oplus\alpha=\text{mod}\left\{ \beta_k + \alpha, 2\pi \right \}$, with $\beta_k=\frac{2\pi}{n}k$. In the OAM basis, Eq.~(\ref{eq:NFracStateSup2}) becomes
%%% equation
\begin{align}
	\ket{M,\alpha}_n =& \mathcal{N} \sum^{n-1}_{k=0} \left\{ \sum_{\ell} c_{\ell, M}(\beta{_k}\oplus\alpha)  \ket{\ell} \right\} \nonumber, \\
	=&\mathcal{N} \sum_{\ell}c^{n}_{\ell, M}(\alpha) \ket{\ell}\,,
\end{align}
%%%%
where the coefficients $c^{n}_{\ell, M}(\alpha)$ are computed from 
%%% equation
\begin{align}
	c^{n}_{\ell, M}(\alpha) = \sum^{n-1}_{k=0} c_{\ell, M}(\beta{_k}\oplus\alpha).
\end{align}
%%%
Using Eq. ~(\ref{eq:fracCoeff}) and the condition $\mod\{\beta_k \oplus \alpha, 2\pi\}=0$, we obtain
%%% equation
\begin{align}
	c^{n}_{\ell, M}(\alpha) =  c_{\ell, M}(\alpha) \sum^{n-1}_{k=0} e^{ i\beta_k\ell }.
\end{align}
%%%
\noindent Since the summation can be evaluated as a geometric series, after some simplification it results in
\[
\sum_{k=0}^{n-1} e^{ i\beta_k\ell } = e^{-i \pi \ell(n-1)/n} \csc\left(\frac{\pi\ell}{n}\right)\sin( \pi \ell ).
\]
Therefore the coefficients can be written as
%%% equation
\begin{align}
	c^{n}_{\ell,M}(\alpha)=& e^{-i \pi \ell(n-1)/n}   A^{n}_{\ell} \  c_{\ell, M}(\alpha)\,, \label{eq:supfracspec}
\end{align}
where
%% equation
\begin{align}
	A^{n}_{\ell} =& \csc\left(\frac{\pi \ell}{n} \right) 
	\sin\left( \pi \ell \right), \nonumber \\
	=& \left\{\begin{array}{cc}
		0 & \text{mod}\left\{ \ell  ,n \right\} \neq 0 \\[2mm]
		1 & \text{mod}\left\{ \ell , n \right\} =0 \\ \end{array} \right..
\end{align}
%%%
\noindent Consequently, the overlap probabilities are
$P_{\ell,n}=|\mathcal{N} \  A^{n}_{\ell}c_{\ell, M}(\alpha)|^2$. Importantly, the probabilities are independent of $\alpha$. Accordingly, the new spectrum has the amplitudes $|c_{\ell, M}|$,  but following the selection rule $ A^{n}_{\ell}$. Indeed, this new spectrum can be tuned by carefully selecting $n$, therefore enabling control of the OAM subspaces.

%%%%%%%%%%%%%%%%%%%%%%%%%%%%%%%%%%%%%%%%%%%%%%%%%%
%%%%%%%%%%%%%%%%%%%%%%%%%%%%%%%%%%%%%%%%%%%%%%%%%%
\section{Spiral imaging of fractional OAM analysers}
Our fractional OAM analysers can be decomposed into the OAM basis using entangled photons through digital spiral imaging. In this scheme, one photon from an entangled pair interacts with the analyser while its twin is decomposed in the OAM basis. The entangled photon pair has a biphoton state 
%%% eq
\begin{equation}\label{eq:Entstate}
	\ket{\Psi}=\sum_{\ell=-L}^L \lambda_{\ell}\ket{\ell} \ket{-\ell},
\end{equation}
%%%
as defined in Eq. ~(\ref{eq: spdc}). The probability amplitude for detecting the  $m$th OAM mode, given a M charged fractional mode of $n$ superpositions, is
%%% eq
\begin{align}
	\tilde{c}^n_m(\alpha) =& \mathcal{M}\bra{m} \braket{M, \alpha |_n |\Psi}, \nonumber \\
	=& \mathcal{M} \ \sum_{\ell=-L}^L \lambda_{\ell}\braket{m|\ell}  \braket{M, \alpha |_n | -\ell}.
\end{align}
%%%
where $\mathcal{M}$ is a normalisation constant such that $\sum_m |\tilde{c}^n_m(\alpha)|^2 =1$. Due to the orthonormality of the OAM basis, the  overlap $\braket{m|\ell}$ is simply the Kronecker delta function $\delta_{m,\ell}$, which evaluates as 0 if $\ell\neq m$ or 1 if $\ell=m$. Since from Eq. ~(\ref{eq:supfracspec}) we know the expansion coefficients for the analyser in terms of the OAM basis, $\braket{M, \alpha |_n |\ell}$ evaluates as
%%% eq
\begin{align}
	\tilde{c} ^n_m(\alpha) = & \mathcal{M} \  \sum_{\ell=-L}^L \delta_{m,\ell} \lambda_{\ell} \ \left[\mathcal{N} c^{n}_{-\ell}(\alpha)\right]^* \nonumber \\ 
	= & \mathcal{M} \mathcal{N} \ \lambda_{m} \ \left[ c^{n}_{-m}(\alpha) \right]^* .
	\label{eq:SpectrumFrac}
\end{align}
%%%
These new weightings are simply the original coefficients of the analysers modulated by the spectrum of the entangled system.~For a maximally entangled state, we obtain the expression $|\tilde{c}^n_m(\alpha)|^2 = |c^{n}_{-m}(\alpha)|^2$, being the original weightings of the analyser, as desired.

\begin{figure}[t]
	\centering
	\includegraphics[width=1\linewidth]{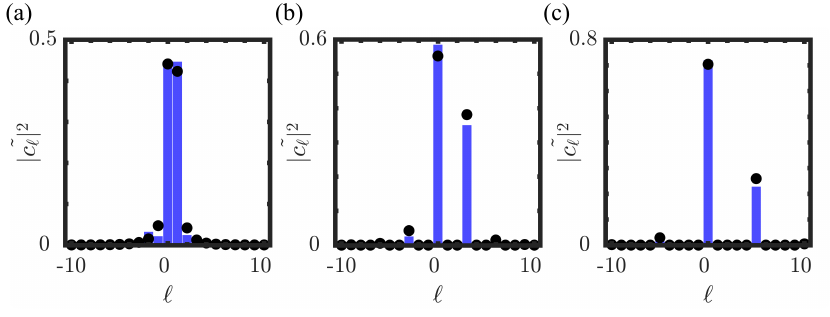}
	\caption{Measured (bars) and theoretical spectrum (points) for fractional OAM analysers  (a) $\ket{M,\alpha}_n=\ket{0.5,0}_1$, (b) $\ket{M,\alpha}_n=\ket{1.5,0}_3$, and (c) $\ket{M,\alpha}_n=\ket{2.5,0}_5$ resulting from digital spiral imaging with entangled photons. Here the weightings are modulated by the OAM spectrum of the entanglement source according to Eq. ~(\ref{eq:SpectrumFrac}).} \label{fig:DecompSP} 
\end{figure}

In Fig.~\ref{fig:DecompSP} we show the measured weightings for our SPDC system which has a normally distribution of OAM modes with $\Delta \ell=11$ centered at $\ell=0$. We show results for $\ket{M,\alpha}_n = \ket{0.5,0}_1,\ket{1.5,0}_3,\ket{2.5,0}_5$ for analysers $n=1,3$ and $5$ in Fig.~\ref{fig:DecompSP}(a),(b) and (c), respectively. It can be seen that the theory (points) and experiment (bars) are in good agreement. To obtain these results, two photons where generated from an SPDC source and modulated with SLMs (see experimental setup in the Methods section in the main text). One SLM was encoded with a fractional OAM mode projecting onto the state, $\ket{M,0}_n$, while the second SLM was encoded with OAM basis modes, $\ket{\ell}$. 
%%% figure
%%%

%%%%%%%%%%%%%%%%%%%%%%%%%%%%%%%%%%%%%%%%%%%%%%%%%%
%%%%%%%%%%%%%%%%%%%%%%%%%%%%%%%%%%%%%%%%%%%%%%%%%%
\section{Detection probability from relative rotations of the analysers}

%%% figure
\begin{figure*}[htbp]
	\centering
	\includegraphics[width=1\linewidth]{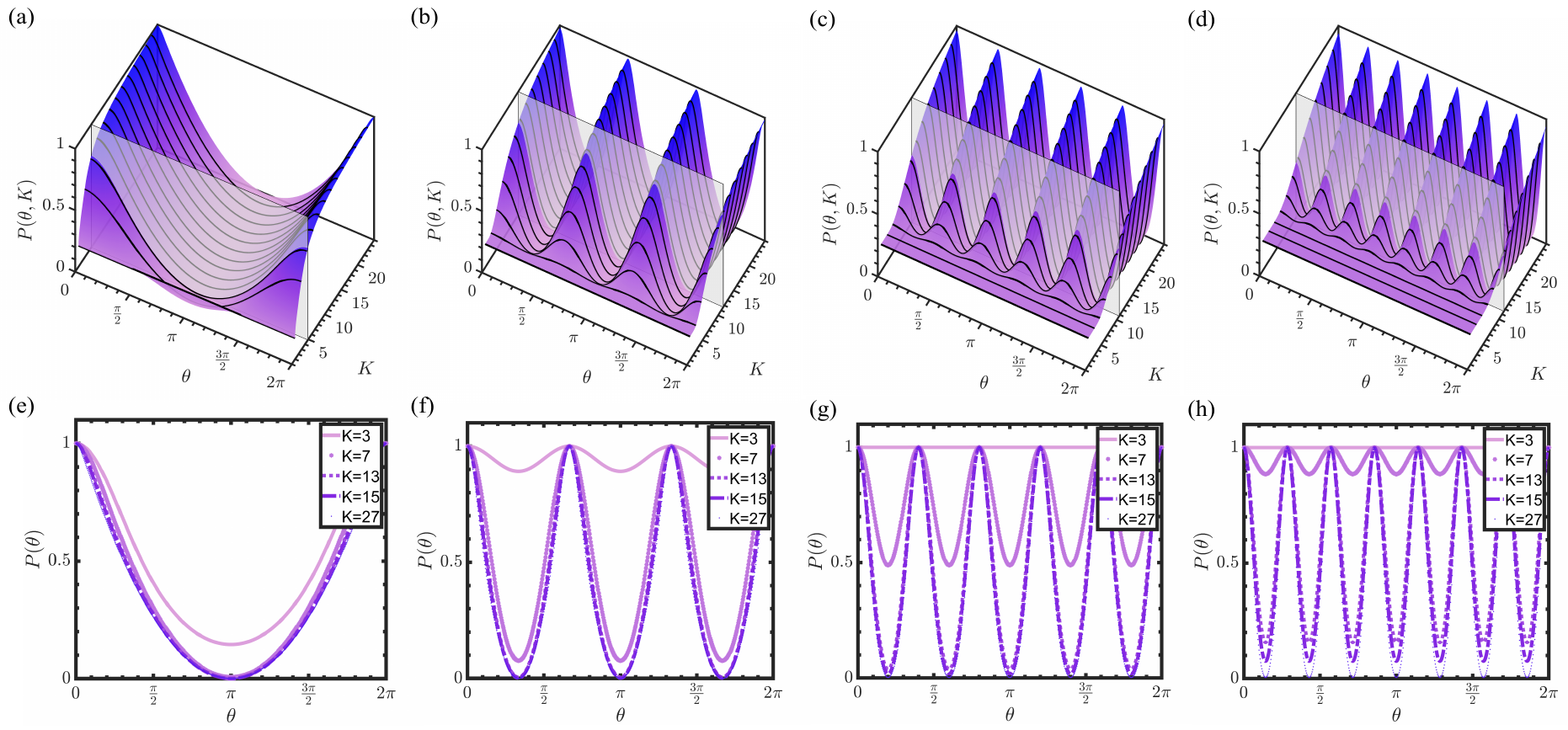}
	\caption{ Simulated normalised probability curves for (a) $n$ = 1, (b) $n$ = 3, (c) $n$ = 5 and (d) $n$ = 7, with fractional OAM $M=\frac{n}{2}$, as a function of the relative orientation $\theta$ between the two analysers and the dimensions, $K$, of the entanglement state. The second row of panels (e)-(h) are probability curves for specific $K$ values for each analyser. The normalisation to unity was performed to illustrate the impact of the dimensions on the visibility. Here, the OAM spectrum shape was assumed to follow a normal (Gaussian) distribution.}\label{fig:Curves_for_K} 
\end{figure*}
%%%

Given a bipartite system of the form of Eq.~(\ref{eq:Entstate}), we want to know what the detection probability is, due to the relative rotations of our fractional OAM analysers acting the entangled photons. Suppose the first analyser  projects onto the state $\ket{M, \theta_1}_n$, and  the second analyser projects on the state $\ket{-M, \theta_2}_n$. A joint measurement on a two photon system using the two analysers is characterized by the product state $\ket{M, \theta_1}_n \ket{-M, \theta_2}_n$. The probability amplitude resulting from such a measurement is
%%% eq
\begin{align}
	C_n(\theta_1, \theta_2) =& \bra{\theta_2, -M}_n \braket{\theta_1, M|_n |\Psi} \nonumber \\
	=& \sum_{\ell=-\infty}^\infty  \lambda_{\ell} \  \braket{ \theta_1, M|_n |\ell } \ \braket{ \theta_2, -M | _n | -\ell}.\label{eq:PampSPDC}
\end{align}
%%%
Therefore we only need to know how to decompose each of the analysers in the OAM basis to obtain the detection probability for the joint measurements. Using Eq.~(\ref{eq:supfracspec}), it follows that
%%% eq
\begin{align} 
	C_n(\theta_1, \theta_2)  \propto &  \sum^{\infty}_{\ell=-\infty}  \underbrace{\lambda_{\ell} }_{\text{SPDC}} \ [ \ \underbrace{c_{\ell, M}^n(\theta_1)}_\text{analyser} 
	\  \underbrace{c_{-\ell, -M}^{n} (\theta_2) }_\text{analyser} \ ]^{*}.
\end{align}
%%%
We use this approach to numerically calculate the detection probabilities $|C_n(\theta_1, \theta_2)|^2$ by simply calculating the probability amplitudes for each analyser in the OAM basis with a desired rotation $\theta_{1,2}$ and multiplying them with the coefficients $\lambda_{\ell}$ that determine the quantum system being probed.

An alternative approach, can be to compute the overlap integral by considering the modal overlaps in the azimuthal degree of freedom, $\phi$, following
\[
\braket{\theta,M|_n |\ell}= \frac{1}{2\pi} \int \exp(-i\Phi_M(\phi;\theta)) \times \exp(i \ell \phi) \ d\phi,
\]
with $\Phi_M(\phi;\theta)/\sqrt{2\pi}$ being the transmission function of the fractional OAM analyser projects onto the state $\ket{M,\theta}_n$. We can rewrite probability amplitude  $C_n(\theta_1, \theta_2)$ as an overlap integral given
%%% eq
\begin{multline}
	C_n(\theta_1, \theta_2)  =  \frac{1}{4\pi} \sum_{\ell=-\infty}^{\infty} \left( \lambda_{\ell}  \iint e^{-i\Phi_M(\phi_1;\theta_1)} e^{i\ell\phi_1} \right.\\
	\left.\vphantom{\iint e^{-i\Phi_M(\phi_1;\theta_1)}}\times e^{-i\Phi_{-M}(\phi_2;\theta_2)} e^{-i\ell\phi_2} \ d\phi_1 d\phi_2 \right),
\end{multline}
%%%
where $\Phi_{\pm M}(\phi_{1,2}, \theta_{1,2})$ are the phases of the fractional OAM analysers. Since $e^{-i\Phi_M(\phi_1;\theta_1)}$ has no $\ell$ dependence, we can introduce the summation into the second integral resulting in
%%% eq
\begin{multline}
	C_n(\theta_1, \theta_2)  =  \frac{1}{2\pi} \int e^{-i\Phi_M(\phi_1;\theta_1)} \ \left(\vphantom{\sum_{\ell=-\infty}^{\infty}} \int  e^{-i\Phi_M(\phi_2;\theta_2)}\right.\\
	\left.\times \frac{1}{2\pi} \sum_{\ell=-\infty}^{\infty} \lambda_{\ell}  e^{i\ell\left(\phi_1 -\phi_2\right)}  d\phi_2\right) d\phi_1.
\end{multline}
%%%
It is convenient to define the periodic function
\[
\Lambda(\phi_1-\phi_2) = \frac{1}{2\pi} \sum_{\ell=-\infty}^{\infty} \lambda_{\ell}  e^{i\ell\left(\phi_1 -\phi_2\right)},
\]
with angular harmonics $e^{i\ell\left(\phi_1 -\phi_2\right)}$ determined by the coefficients $\lambda_{\ell}$, and use it to rewrite $C_n(\theta_1, \theta_2)$ as
%%% eq
\begin{multline}
	C_n(\theta_1, \theta_2) =  \frac{1}{2\pi} \int e^{-i\Phi_M(\phi_1;\theta_1)} \\ \left( \int  e^{-i\Phi_M(\phi_2;\theta_2)} \Lambda(\phi_1 -\phi_2)d\phi_2\right) d\phi_1.
\end{multline}
%%%
Notice that the second integral is a convolution between $\Lambda(\phi_1-\phi_2)$ and the second analyser. As a example, we consider a maximally entangled state ($\lambda_\ell:=\text{constant}$). In this case, $\Lambda(\phi_1 -\phi_2)=\delta(\phi_1 -\phi_2)$ and therefore
%%% eq
\[
C_n(\theta_1, \theta_2)  =  \frac{1}{2\pi}\int e^{-i\Phi_M(\phi;\theta_1)} \ e^{-i\Phi_{-M}(\phi;\theta_2)}  d\phi.
\]
%%%
The integral now only depends in the transmission functions of the analysers with an analytical solution found in \cite{huang2018various}.

We now calculate the probability $P_n(\theta_1, \theta_2)=\left|C_n(\theta_1, \theta_2)\right|^2$ as a function of relative orientation  $\theta=(\theta_1-\theta_2)$ between the two analysers and the dimensions, $K$, of an entangled system with some given OAM spectrum $|\lambda_\ell|^2$. The latter is embedded in the function $\Lambda(\phi_1 -\phi_2)$. Figures~\ref{fig:Curves_for_K}(a)-(d) show examples of the probability surfaces assuming a normal (Gaussian) spectrum $|\lambda_\ell|^2$ for superposition states $n= 1, 3, 5$ and $7$. In the second row of, Fig.~\ref{fig:Curves_for_K} (e)-(h), we show examples of the probability curves normalised to unity for several values of dimensionality $K$. Here, it can be seen that the frequency of the probabilities as a function of $\theta$ increases with $n$, owing to the n-fold symmetry in the phase profiles of the analysers.

\begin{figure*}[t]
	\centering
	\includegraphics[width=0.9\linewidth]{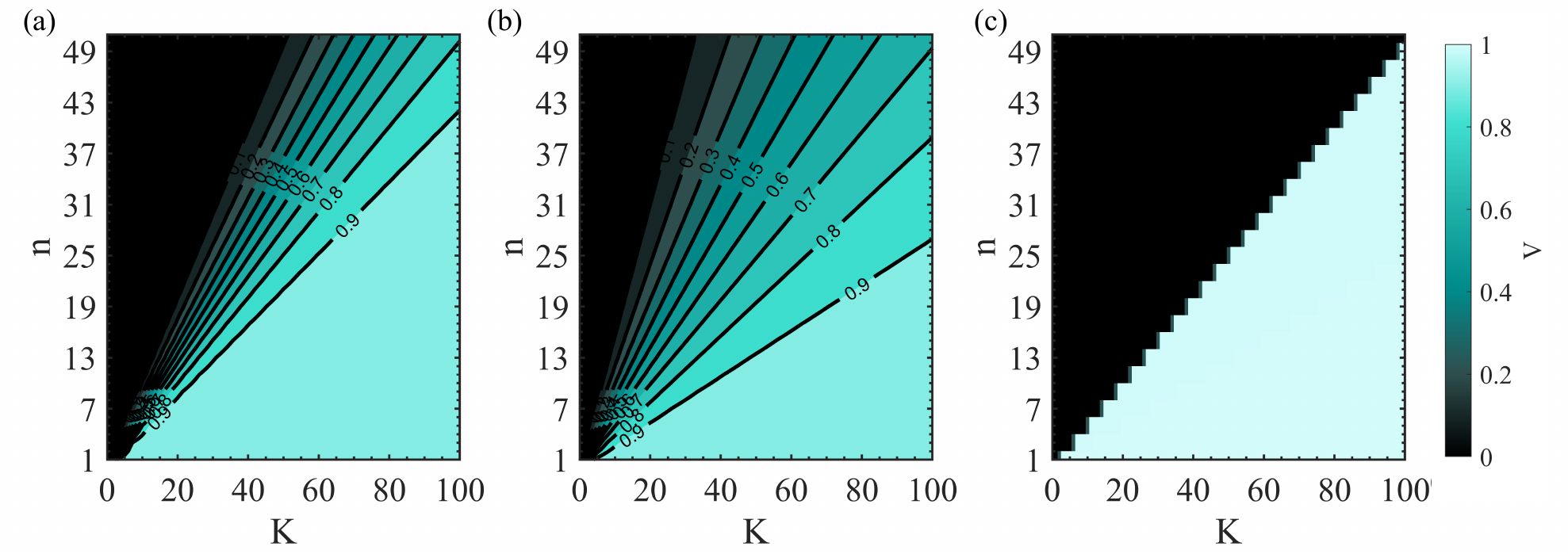}
	\caption{ Contour plots of visibility vs dimensionality ($K$)  vs $n$ (number of fractional OAM superpositions) for the (a) Normal, (b) SPDC theory and (c) maximally uniform distribution ( or maximally entangled pure state). Here we demonstrate the sensitivity of the analysers to the dimensions of a OAM entanglement. The visibilities from the maximally entangled state demonstrates the minimum number of modes required to have a visibility $V=1$. }\label{fig:VvsK}
\end{figure*}

%%% figure
\begin{figure*}[t]
	\centering
	\includegraphics[width=1\linewidth]{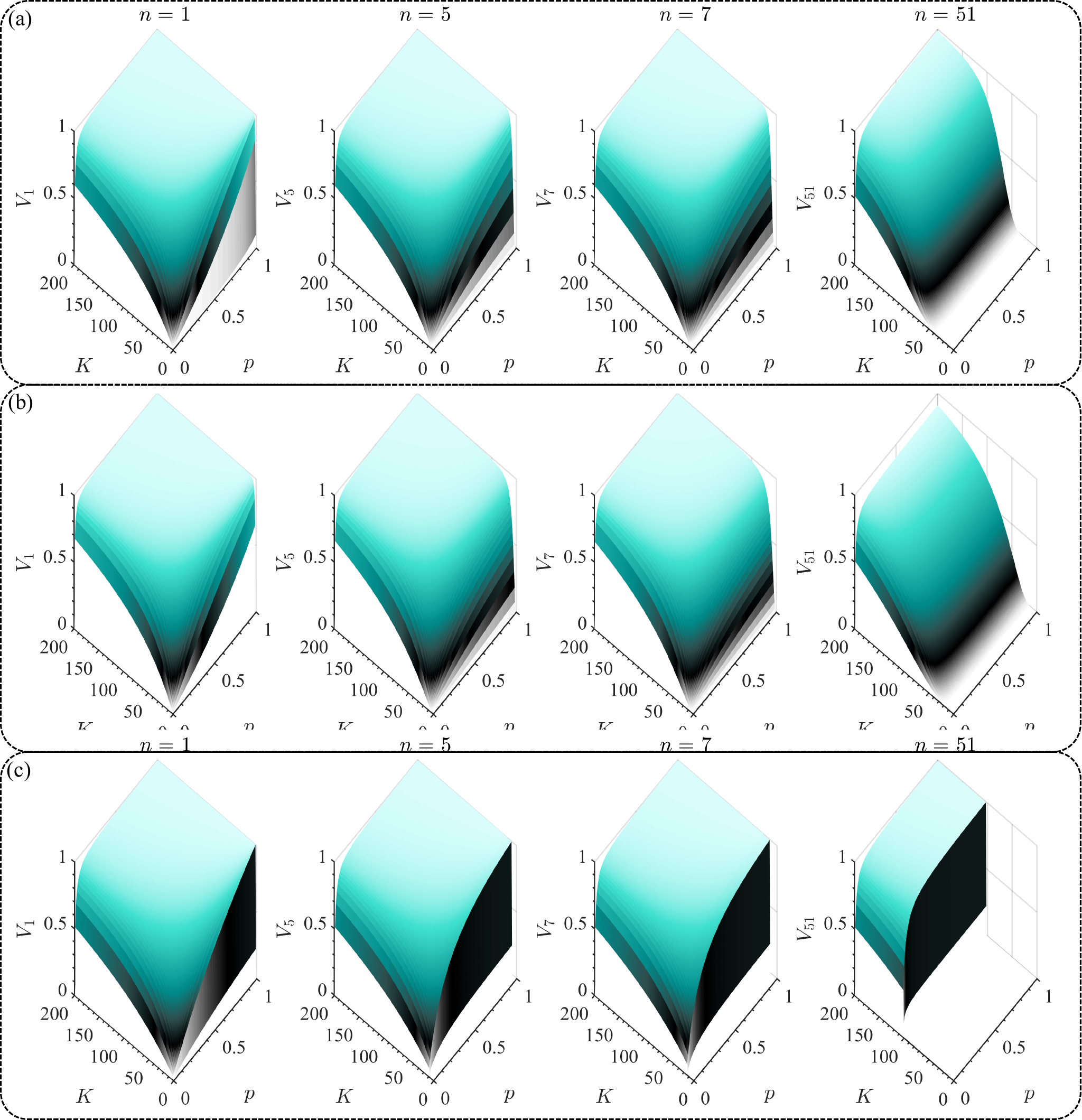}
	\caption{Visibility as a function of purity (p) and dimensions  (K) for the (a) Normal, (b) SPDC theory and (c) the uniform (maximally entangled state) obtained for fractional OAM projections corresponding to  $n=1,5,7,51$. }\label{fig:VvsKvsp}
\end{figure*}
%%%

Crucially, the exact shape and visibility of the curves depends on both the dimensions ($K$) of the state being probed and the number of superpositions ($n$). For all $n$'s, the visibility for a specific $K$ shows a decreasing trend as the number of superpositions $n$ are increased. Therefore the analysers are sensitive to the dimensions of the system.

We also found that the shape of the spectrum affects the measured probabilities, as illustrated in Fig.~\ref{fig:VvsK}(a)-(d) for square (maximally entangled), normal (Gaussian) and SPDC, respectively.

Now that we have shown how the detected probabilities depended on the dimensions and superposition states measured, in the following section we study the relation between the visibility and dimensions quantitatively.

%%%%%%%%%%%%%%%%%%%%%%%%%%%%%%%%%%%%%%%%%%%%%%%%%%
%%%%%%%%%%%%%%%%%%%%%%%%%%%%%%%%%%%%%%%%%%%%%%%%%%
\section{Visibility for different spectra}

The visibilities are calculated from detection probabilities resulting from the projections of an entangled state with an initial OAM distribution $|\lambda_\ell|^2$  onto the states $\ket{M, 0}_{n}\ket{-M,\theta}_{n}$, where $\theta\in[0, 2\pi]$ is their relative rotation.

For example, for a square (uniform) ($K\rightarrow \infty$) distribution and $n$ superpositions of fractional modes the probability is given by \cite{huang2018various}
%%% eq
\begin{align}
	P(\theta_1, \theta_2) =&  |C(\theta_1, \theta_2)|^2 \nonumber\\
	=& a \sin^2\left(\frac{M\pi}{n}\right)+\cos^2\left(\frac{M\pi}{n}\right),
\end{align}
%%%
\noindent  with $a = (\pi(2t-1)-n\theta)^2/\pi^2$ for $\frac{2\pi}{n}(t-1) \leq \theta\leq\frac{2\pi}{n}(t), \quad t=1,...,n,$ where $t$ indexes each $2\pi/n$ period over the range of $0\leq\theta<2\pi$ and $\theta=\theta_1-\theta_2$. This oscillating function results in fringes with a visibility function given by
%%% eq
\[
V_n(M)=\frac{1 - \cos^{2}\left( \frac{M\pi}{n} \right)}{1+\cos^{2}\left( \frac{M\pi}{n} \right)} .
\]
%%%
For $n=1$, parabolic fringes with perfect visibility occur when $M=\ell+0.5$ for all OAM integer charges $\ell$. In contrast, when $n>1$ high visibility fringes occur for only specific choices of $n$ and $M$. That is, parabolic fringes with high visibility ($V=1$) are expected when $n$ is odd and $\text{mod}\left\{M-\frac{n}{2},n\right\}=0$.

Contour plots of the visibilities with changing dimensions ($K$) and fractional OAM superpositions ($n$) for various OAM spectral shapes (Normal, SPDC, Uniform) are shown in Fig. \ref{fig:VvsK} (a)-(c) for pure states. As shown, the assumed spectrum can affect the visibility that is measured for various superpositions ($n$). The visibilities for each analyser ($n$) and spectrum shape are monotonic with increasing $K$. In particular, for the uniform spectrum (maximal entanglement in K dimensions) the visibility is 1 above some $K=d_n$ and zero below this. We further exploit this property to determine the dimensionality of an entanglement system. 
%%%%%%%%%%%%%%%%%%%%%%%%%%%%%%%%%%%%%%%%%%%%%%%%%%
%%%%%%%%%%%%%%%%%%%%%%%%%%%%%%%%%%%%%%%%%%%%%%%%%%
\section{Visibility of mixed states}

The visibilities that can be measured with our analysers are not only dependent on the effective dimensions of the system but also the purity. In particular, we consider the isotropic state, 

%%%
\begin{equation}
	\rho_p = p \ket{\Psi_d}\bra{\Psi_d} + \frac{1-p}{d^2}\mathbb{I}_{d^2},
\end{equation}
which can decomposed into the high-dimensional entangled state, $\ket{\Psi_d}$, and the separable and mixed state, $1/d^2\mathbb{I}_{d^2} =1/d^2 \sum_{\ell, \ell' =-L} ^{\ell,\ell'=L} \ket{\ell}\ket{\ell'}\bra{\ell'}\bra{\ell}$, where $\mathbb{I}_{d^2}$ is the identity operator. Such states model quantum systems that have noise contributions from the environment. Here $p$ can be associated with the  purity of the state ranging from a maximally mixed ($p=0$) to a pure state ($p=1$). Interestingly, the isotropic state is separable for $p \leq 1/(d+1)$ and entangled otherwise. Importantly the generalised Bell inequality can also be violated when $p>2/S_d$ where $S_d$ is the Bell parameter \cite{Collins2002}. We show that both $p$ and $d$ can be measured using our analysers. For convenience, we assume $d \approx K$, where K is the effective dimensionality of the entanglement. We will demonstrate that we can measure both $p$ and $K$ using our analysers.

Firstly, we calculate the detection probabilities from the overlap, $P_{n}(\theta; K, p) = \text{Trace}(\hat{M}\rho_p)$ where $\hat{M}$ projects onto the states $\ket{M,0}_n\ket{-M, \theta}_n$. As a result, the detection probability can be written as
\begin{equation}
	P_n(\theta; p, K)  = p P_n(\theta; K) +  \frac{1-p}{K^2} I_n(0; K),
\end{equation}
where $P_n(\theta, K)=\left|\sum_{\ell=-L}^{L} \lambda_{\ell} c^{n}_\ell(0)c^{n}_{-\ell}(\theta)\right|^2$ and $I_n(0; K) =\left|\sum_{\ell=-L}^{L} \left|c^{n}_\ell(0)\right|^2 \ \right|^2$ is the overlap of the analysers with the maximally mixed state. Since the functions are periodic and obtain maximum and minimum values for $\theta = 0$ and $\pi/n$, respectively, we obtain the expression
\begin{align}
	\Delta P_n( p, K)  & = P_n(0; p, K) - P_n(\pi/n; p, K) \nonumber\\
	& = p \Delta P_n(K),
\end{align}
where $\Delta  P_n(K) = P_n(0,p=1, K)- P_n(\pi/n,p=1,K)$. The visibilities can be calculated from
\begin{align}
	V_n(p,K)&=\frac{\Delta P_n( p, K)}{P_n(\pi/n;p, K) + P_n(0;p,K)},
\end{align}
We show the dependence of the visibilities on the dimensions ($K$) and purity ($p$) in Fig. \ref{fig:VvsKvsp} (a-c) for the Normal, SPDC and uniform distribution, respectively. Each panel shows the visibilties from various analysers depending on the number of superposition ($n$). As shown the visibilities increase monotonically with increasing dimensions ($K$) as well as purity $p$ for each analyser where $p=1$ obtains a maximal visibility. However, as $n$ increases the visibilities  decrease for all $p$ and $K$. Since the visibilities are monotonic in both $p$ and $K$ as well as $n$, we can exploit this property to map the dimensions of a quantum system. We favour this approach since the visibilities can be easily measured and  require few measurements (peak and trough). 

\section{Comparison between a known and guessed spectrum}
\begin{table*}[h]
	\begin{tabular}{|c|c|c|c|c|c|c|c|} 
		\hline
		Noise level & $p^{SPDC}$ & $K^{SPDC}$ & $p^\text{normal}$ & $K^\text{normal}$ & $Q$ & $\hat{K}$ & $\hat{p}$\\
		\hline
		low     &0.45 $\pm$ 0.03    &22.84 $\pm$ 0.62  &0.42 $\pm$ 0.02    &20.00 $\pm$ 0.32     &  19.19 & 22 & 0.44 \\ \hline
		% medium  &0.20 $\pm$ 0.02    &25.90 $\pm$ 1.12  &0.23 $\pm$ 0.01    &21.72$\pm$ 0.03   & 10.25 & 25 & 0.27\\ \hline
		high    &0.13 $\pm$ 0.01    &17.73 $\pm$ 0.71  &0.13 $\pm$ 0.01    &17.18  $\pm$ 0.34   & 3.76 & 18 & 0.13\\ \hline
	\end{tabular}
	\caption{Measured purity ($p$) and dimensionality ($K$), under low and high noise levels, compared to estimates from other methods. Here $Q$ is the average quantum contrast.}
	\label{tab:purity_dimen_summary}
\end{table*}
Using our procedure we measured the dimensions and purity of SPDC photons with varying noise levels (low and high).  The results are summarised in Table \ref{tab:purity_dimen_summary}. In the second and third column, we know what the input spectrum shape (SPDC) is and can therefore accurately optimise for the dimensions ($K$) and purity ($p$) of the state (see Results section). Further, if we guess the spectrum based on its shape (symmetry) we also obtain values that are similar to the expected results, with a relative error of up to $\approx13\%$. This was done using the normal distribution as the function modelling the mode spectrum. Next, we verify  our result using the values extracted from the spiral bandwidth.

To calculate the the expected dimensions, $\hat{K}$, we used the coincidences from the spiral spectrum in the OAM basis, i.e $C_{\ell_A, m_B}$, where $\ell_A$ denotes the mode index of photon A and $m_B$ for photon B. Since we want the Schmidt number of the pure part of the state, we subtracted the accidentals and then used Eq. (\ref{eq: scmidt}), yielding results with a low relative error of $3\%$, validating our results. Subsequently, we estimated the purity $\hat{p}$ (see Results section). Note that no accidental subtraction was performed in this case. Accordingly, to estimate the purity we measured the quantum contrast using
\begin{equation}
	Q =  \Bar{C} /  \Bar{C}',
\end{equation}
taken from the ratio between the average coincidences in the anti-diagonal entries, $\Bar{C} = \sum_\ell C_{\ell, -\ell}$ and the average noise contribution from coincidences excluding the anti-diagonal entries, i.e  $\Bar{C}' = \frac{1}{d^2-d} \left( C_T -d \ \Bar{C} \right)$. Here $C_T$ corresponds to the total coincidences $C_T=\sum_{\ell , m} C_{\ell, m}$. Indeed, using the quantum contrast we obtained a purity that is comparable to that obtained from our method showing a relative error of only up to $2\%$.

% %%% figure
% \begin{figure}[t]
% 	\centering
% 	\includegraphics[width=0.95\linewidth]{ChiPplot.pdf}
% 	\caption{A plot of $1/ \chi(p, K)$ for measured data assuming a Gaussian spectrum for the OAM correlations. The peak represents the optimal purity, $p$, and effective dimensions $K$, that best fits our measured data.}\label{fig:Chiplot}
% \end{figure}
% %%%
% %%% figure
% \begin{figure*}[h]
% 	\centering
% 	\includegraphics[width=0.9\linewidth]{N8ErrorPlot.pdf}
% 	\caption{ The simulated error percentage for various known inputs of purity, $p$ (a)-(c),  and dimensions $K$ (d)-(e). The number of measurement analysers ranges from $N=4, 8$ to, $16$ from the left to the right. The accuracy of the measurement increases with $N$.}\label{fig:accuracytest}
% \end{figure*}

%%%
%\bibliography{mybibfile}

\end{document}